\documentclass[aps,prb,floats,showpacs,twocolumn]{revtex4-1}
\usepackage{graphicx,color}
\begin{document}
\title{Fictitious gauge fields in bilayer graphene}
\author{Eros Mariani$^1$, Alexander J. Pearce$^1$ and Felix von Oppen$^2$}
\affiliation{$^1$ Centre for Graphene Science, School of Physics, University of Exeter, Stocker Rd., EX4 4QL Exeter, UK\\
$^2$Dahlem Center for Complex Quantum Systems and Fachbereich Physik, Freie Universit\"at Berlin, 14195 Berlin, Germany}
\date{\today}
\begin{abstract}
We discuss the effect of elastic deformations on the electronic properties of bilayer graphene membranes. Distortions of the lattice translate into fictitious gauge fields in the electronic Dirac Hamiltonian which are explicitly derived here for arbitrary elastic deformations. We include gauge fields associated to intra- as well as inter-layer hopping terms and discuss their effects in different contexts. As a first application, we use the gauge fields in order to study the recently predicted strain-induced Lifshitz transition for the Fermi surface at low energy.
As a second application, we discuss the electron-phonon coupling induced by the fictitious gauge fields and analyse its contribution to the electrical resistivity of suspended bilayer membranes. Of special interest is the appearance of a linear coupling for flexural modes, in stark contrast to the case of monolayer graphene. This new coupling channel is shown to dominate the temperature-dependent resistivity in suspended samples with low tension.
\end{abstract}
\maketitle

\section{Introduction}
Graphene, a monolayer of carbon atoms, is the only two-dimensional (2D) conducting elastic membrane, where electrons have been shown to behave as massless Dirac fermions.\cite{Geim,Kim} Since the experimental realisation of graphene, several authors investigated the interplay between its electronic properties and externally induced mechanical deformations (see, e.g. Ref.\ \onlinecite{Vozmediano} and references therein). This question acquires a particular interest since elastic distortions of the honeycomb lattice translate into fictitious gauge fields in the electronic Dirac Hamiltonian.\cite{Vozmediano,Mahan,Suzuura,Mariani08} Indeed, in presence of deformations, the low-energy massless Dirac Hamiltonian of electrons in monolayer graphene is described by $H=v\mbox{\boldmath$\sigma$}\cdot\left({\bf p}+e{\bf A}\right)$ with the fictitious vector potential ${\bf A}\propto \left(u_{xx}^{}-u_{yy}^{},-2u_{xy}^{}\right)$,\cite{Vozmediano,Mahan,Suzuura,Mariani08} expressed in terms of the strain tensor $u_{ij}^{}$ of the 2D membrane. The latter reflects the symmetry of the system with respect to the plane of graphene, as it is linear in the in-plane deformations and quadratic in the out-of-plane (flexural) distortions. As a direct consequence of this vector potential, it has recently been observed experimentally that strained graphene bubbles exhibit the formation of Landau levels corresponding to magnetic fields as large as $300\; {\rm T}$.\cite{Crommie} This adds significant motivation to recent works devoted to the investigation of the so-called strain-engineering, i.e. the tailoring of the electronic properties of graphene via controllable elastic deformations.\cite{PacoNat,Psemagnanorib}

Bilayer graphene membranes show an even richer scenario which is attracting ever growing interest. This is partly due to their larger potential for device applications, since a gap in the spectrum can be induced by external electric fields.\cite{Gap,McCann,Nilsson,Oostinga,Lau} From a fundamental point of view, the low-energy electronic band structure of bilayer graphene shows fascinating structures, such as the appearance of four massless Dirac cones which evolve into a massive quasiparticle spectrum at higher energy.\cite{McCann} Given the remarkable tunability of the electron density in graphene devices, this opens the possibility of observing a phase transition in the topology of the Fermi surface (FS) as a function of the electron doping, the Lifshitz transition (LT).\cite{Lifshitz,Korean,Falko11} 

So far, the investigation of the effects of elastic deformations in bilayer graphene has been limited to uniaxial deformations.\cite{Korean,Falko11} In particular, the form of the fictitious gauge fields associated to arbitrary distortions is still unknown. These fields can result in dramatic effects on the electronic band structure, and are the starting point for any systematic investigation of electron-phonon coupling and of strain engineering in bilayer samples. In this paper we address this open question by deriving the form of the gauge fields and discussing how they affect the electronic Dirac Hamiltonian. 

As a first application of our analysis we consider the special case of elastic deformations yielding uniform fictitious gauge fields (like uniaxial strain, pure shear, or a rigid shift between the two layers), including some that have been recently discussed in the literature. \cite{Korean,Falko11} These deformations dramatically modify the electronic bands at low energy and induce the annihilation of two massless cones beyond a critical value of strain. This affects the nature of the LT and the energy at which it occurs, with observable consequences on the formation of Landau levels (LL) under perpendicular magnetic fields and on the single particle density of states (DOS) at the Fermi level. The results of this section reproduce those that appeared recently \cite{Korean,Falko11} and serve as a test for the general gauge fields we deduce. 

As a further application of the fictitious gauge fields, we discuss the contribution to the electronic resistivity of suspended bilayer membranes due to electron-phonon coupling. In contrast with the case of monolayers, the two layers are not equivalent and the system does not exhibit a symmetry with respect to out-of-plane deformations. This is reflected in the appearance of a linear intrinsic coupling for flexural phonons in bilayer graphene, in contrast to the purely quadratic one present in monolayers.\cite{Mariani08,Mariani10,Castro,Ochoa10} This linear coupling channel results in the dominant electron-phonon contribution to the resistivity for suspended samples with low tension.  
Our investigation provides the basis for the microscopic study of the electro-mechanical properties of bilayer graphene membranes, including the fictitious magnetic fields resulting from arbitrary elastic mechanical deformations.

The structure of the paper is as follows: In Sec.\ \ref{sec:Basics} we briefly summarise the electronic properties of bilayer graphene, considering all hopping parameters in the original tight-binding Hamiltonian. In Sec.\ \ref{sec:Gauge} we derive the explicit expression for the fictitious gauge fields in bilayer graphene under arbitrary elastic deformations, and deduce the effective low-energy Hamiltonian for the two bands close to zero energy. Section \ref{sec:NoStrain} summarises the electronic band structure in the absence of deformations. In Sec.\ \ref{sec:Strain} we introduce elastic deformations resulting in uniform fictitious gauge fields and discuss their effect on the electronic low-energy spectrum. We analyse the displacement of the massless Dirac cones and their annihilation under a critical strain. As a consequence, we explore the tailoring of the band-structure via controllable elastic deformations, finally resulting in the possibility of inducing topological LT at the Fermi level. In Sec.\ \ref{sec:Resistivity} we discuss the electron-phonon coupling induced by the fictitious gauge fields in suspended bilayer membranes and analyse its contribution to the temperature-dependent component of the resistivity. Finally, in Sec.\ \ref{sec:Conclusions} we present our conclusions.

\section{Electrons in bilayer graphene}
\label{sec:Basics}

\begin{figure}[ht]
	\centering
		\includegraphics[width=0.8\columnwidth]{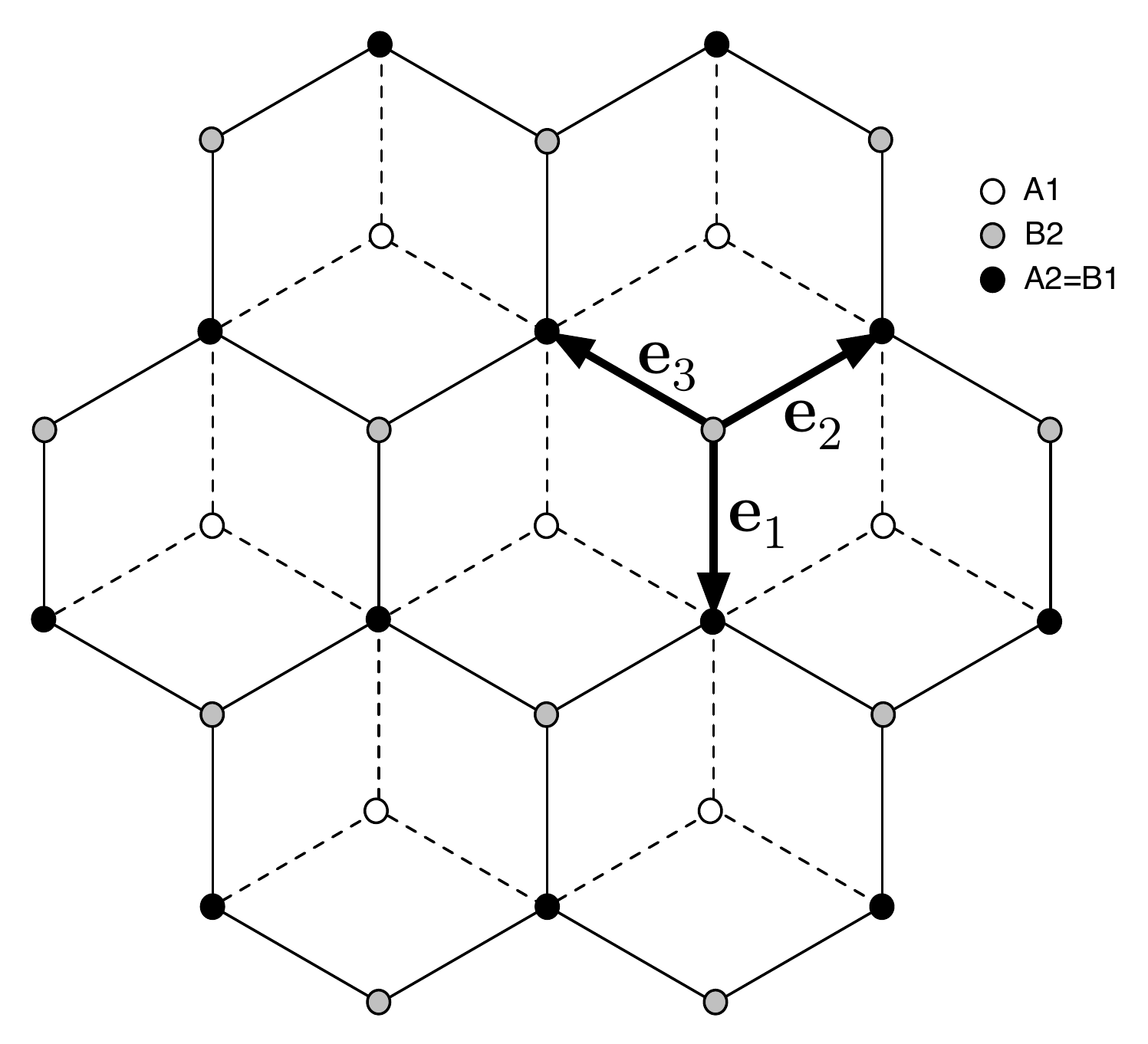}
			\caption{Atomic structure of perfect bilayer graphene in Bernal stacking (top view). The honeycomb lattice with full (dashed) lines corresponds to the upper (lower) layer $l=2$ ($l=1$). The atomic sites $A2$ and $B1$ coincide once projected on the plane. The three vectors ${\bf e}_{j}^{}$, ($j=1,2,3$) connecting $A$ and $B$ sites are indicated (see text for more details).\label{BilayerTopView}}
\end{figure}

The electronic properties of ideal bilayer graphene have been studied in the past by several authors.\cite{McCann,Nilsson} The carbon atoms in the two layers are identified by a layer index ($l=1,2$, for the first and second layer, respectively) and a sublattice index ($s=A,B$), to distinguish inequivalent atomic sites. In each layer, carbon atoms form honeycomb lattices with nearest neighbour distance $a=1.42\; {\rm \AA}$. The three nearest neighbours of type A are obtained from a central B atom by the displacement vectors ${\bf e}_1=a\, (0,-1)$, ${\bf e}_2=a\, (\sqrt{3}/2,1/2)$ and ${\bf e}_3=a\, (-\sqrt{3}/2,1/2)$. 
In the Bernal stacking configuration for a bilayer in the x-y plane, the atoms in layer 2 and sublattice A (i.e. of type A2) are located directly above those of type B1, at a distance $c\simeq 3.34\, {\rm \AA}$, as indicated in Fig.\ \ref{BilayerTopView}.
\\
On the contrary, B2 atoms are not located above A1 ones. In fact, the pairs A1-A2, B1-B2 and A1-B2 are all separated by the distance $\tilde{c}=\sqrt{c^2+a^2}\simeq 3.63\, {\rm \AA}$.
The starting point for the investigation of the electronic properties of bilayer graphene is the tight-binding Hamiltonian in the real space
\begin{eqnarray}
&&H_{{\rm TB}}^{}=-t_{A1,B1}^{}\sum_{{\bf R}_{B1}^{}}^{}\sum_{j=1}^{3}|{\bf R}_{B1}^{}\rangle\langle {\bf R}_{B1}^{}+{\bf e}_{j}^{}|\nonumber\\
&&-t_{A2,B2}^{}\sum_{{\bf R}_{B2}^{}}^{}\sum_{j=1}^{3}|{\bf R}_{B2}^{}\rangle\langle {\bf R}_{B2}^{}+{\bf e}_{j}^{}|\nonumber\\
&&-t_{A2,B1}^{}\sum_{{\bf R}_{B1}^{}}^{}|{\bf R}_{B1}^{}\rangle\langle {\bf R}_{B1}^{}+c\hat{\bf z}|\\
&&-t_{A1,B2}^{} \sum_{{\bf R}_{A1}^{}}^{}\sum_{j=1}^{3}|{\bf R}_{A1}^{}\rangle\langle {\bf R}_{A1}^{}+{\bf e}_{j}^{}+c\hat{\bf z}|\nonumber\\
&&-t_{A1,A2}^{} \sum_{{\bf R}_{A2}^{}}^{}\sum_{j=1}^{3}|{\bf R}_{A2}^{}\rangle\langle {\bf R}_{A2}^{}+{\bf e}_{j}^{}-c\hat{\bf z}|\nonumber\\
&&-t_{B1,B2}^{} \sum_{{\bf R}_{B2}^{}}^{}\sum_{j=1}^{3}|{\bf R}_{B2}^{}\rangle\langle {\bf R}_{B2}^{}+{\bf e}_{j}^{}-c\hat{\bf z}|+h.c.\nonumber
\end{eqnarray}
where ${\bf R}_{sl}^{}$ is the position of an atom in sublattice $s$ and layer $l$, and $|{\bf R}_{sl}^{}\rangle$ is the ket associated to the corresponding localised orbital. For the interlayer hopping terms we used the identities ${\bf R}_{A2}^{}={\bf R}_{B1}^{}+c\hat{\bf z}$, ${\bf R}_{B2}^{}={\bf R}_{A1}^{}+{\bf e}_{j}^{}+c\hat{\bf z}$, ${\bf R}_{A1}^{}={\bf R}_{A2}^{}+{\bf e}_{j}^{}-c\hat{\bf z}$ and ${\bf R}_{B1}^{}={\bf R}_{B2}^{}+{\bf e}_{j}^{}-c\hat{\bf z}$. In the tight-binding Hamiltonian, $t_{sl,s'l'}^{}$ represents the hopping energy between the two neighboring sites at ${\bf R}_{sl}^{}$ and ${\bf R}_{s'l'}^{}$.
In bilayer graphene samples these hopping terms are given by $t_{Aj,Bj}^{}\equiv t_{j}^{}\simeq 2.47\, {\rm eV}$ (with $j=1,2$),\cite{tdata} $t_{A2,B1}^{}\equiv \gamma\simeq 0.39\, {\rm eV}$, \cite{gamma1data} $t_{A1,B2}^{}\equiv \gamma_{3}^{}\simeq 0.315\, {\rm eV}$,\cite{gamma3data} and $t_{A1,A2}^{}=t_{B1,B2}^{}\equiv \gamma_{4}^{}\simeq 0.044\, {\rm eV}$.\cite{gamma4data} For the sake of generality, in the Hamiltonian above we consider the case in which the intralayer hopping energies $t_1^{}$ and $t_2^{}$ can be different. This can be of relevance for bilayers on a substrate, where the direct contact with the latter may affect the hopping energy in one layer with respect to the other. 
Electrons in the bilayer lattice are described by Bloch states of the form 
\begin{eqnarray}
&&|\psi_{{\bf k}}^{}\rangle =\sum_{{\bf R}_{A1}^{}}^{}u^{(A1)}_{{\bf k}}e^{i{\bf k}\cdot {\bf R}_{A1}^{}}|{\bf R}_{A1}^{}\rangle+\sum_{{\bf R}_{B1}^{}}^{}u^{(B1)}_{{\bf k}}e^{i{\bf k}\cdot {\bf R}_{B1}^{}}|{\bf R}_{B1}^{}\rangle \nonumber\\
&&\; +\sum_{{\bf R}_{A2}^{}}^{}u^{(A2)}_{{\bf k}}e^{i{\bf k}\cdot {\bf R}_{A2}^{}}|{\bf R}_{A2}^{}\rangle+\sum_{{\bf R}_{B2}^{}}^{}u^{(B2)}_{{\bf k}}e^{i{\bf k}\cdot {\bf R}_{B2}^{}}|{\bf R}_{B2}^{}\rangle\; ,
\end{eqnarray}
with $u^{(sl)}_{{\bf k}}$ the amplitude of the wavefunction on the sublattice $s$ and layer $l$ at wavevector ${\bf k}$. 
In the $4\times 4$ space of the Bloch amplitudes $(u^{(A1)}_{{\bf k}},u^{(B2)}_{{\bf k}},u^{(A2)}_{{\bf k}},u^{(B1)}_{{\bf k}})$ the Hamiltonian thus takes the form
\begin{equation}
\label{H0}
H_0^{}=\left(
\begin{array}{cccc}
0 & -\gamma_{3}^{} f_{{\bf k}}^{} & -\gamma_{4}^{} f_{{\bf k}}^{*} & -t_1 f_{{\bf k}}^{*}\\
-\gamma_{3}^{} f_{{\bf k}}^{*} & 0 & -t_2 f_{{\bf k}}^{} & -\gamma_{4}^{} f_{{\bf k}}^{}\\
-\gamma_{4}^{} f_{{\bf k}}^{} & -t_2 f_{{\bf k}}^{*} & 0 & -\gamma\\
-t_1 f_{{\bf k}}^{} & -\gamma_{4}^{} f_{{\bf k}}^{*} & -\gamma & 0
\end{array}
\right)\; ,
\end{equation}
with $f_{{\bf k}}^{}=\sum_{j=1}^{3}\exp [i{\bf k}\cdot{\bf e}_{j}^{}]$, resulting in four energy bands. Two of them are at high energy of order 
$\pm \gamma$, while the other two touch close to zero energy, in the vicinity of two Dirac points in the first Brillouin zone given by ${\bf K}_{\pm}^{}$, with ${\bf K}_{\tau}^{}=\tau (4\pi/3\sqrt{3}a,0)$, where $f_{{\bf K}_{\tau}^{}}^{}=0$.
In the vicinity of the two Dirac points, the matrix Hamiltonian in Eq.\ (\ref{H0}) is expanded as 
\begin{eqnarray}
\label{H0+}
&&H_{0}^{(+)}=\left(
\begin{array}{cccc}
0 & v_3^{}p & v_{4}^{}p^{\dagger}_{} & v_1 p^{\dagger}_{}\\
v_3^{}p^{\dagger}_{} & 0 & v_2 p & v_{4}^{} p\\
v_{4}^{} p & v_2 p^{\dagger}_{} & 0 & -\gamma\\
v_1 p & v_{4}^{} p^{\dagger}_{} & -\gamma & 0
\end{array}
\right)\; \\
&&H_{0}^{(-)}=\left(
\begin{array}{cccc}
0 & -v_3^{}p^{\dagger}_{} & -v_4 p & -v_1 p\\
-v_3^{}p & 0 & -v_2 p^{\dagger}_{} & -v_4 p^{\dagger}_{}\\
-v_4 p^{\dagger}_{} & -v_2 p & 0 & -\gamma\\
-v_1 p^{\dagger}_{} & -v_4 p & -\gamma & 0
\end{array}
\right)\; ,\nonumber
\end{eqnarray}
where $p=p_{x}^{}+i p_{y}^{}$ is the complex representation of the two dimensional momentum relative to the Dirac point, $v_{j}^{}=3at_{j}^{}/2\hbar$ for $j=1,2$ and $v_{j}^{}=3a\gamma_{j}^{}/2\hbar$ for $j=3,4$.
The two valleys are thus related by the symmetry $H_{0}^{(-)}=H_{0}^{(+)*}\big|_{p^{}_{}\rightarrow -p^{}_{}}^{}$. We now introduce elastic deformations in the bilayer lattice and derive the consequent fictitious gauge fields in the Dirac Hamiltonian formalism.

\section{Fictitious gauge fields in bilayer graphene}
\label{sec:Gauge}

A generic elastic deformation in the bilayer graphene membrane induces a displacement of the atomic positions which results in the modification of bond lengths between neighboring atoms. The corresponding change in the hopping energies results in corrections to the matrix elements of the $4\times 4$ Hamiltonian $H_{0}^{(\tau )}$, yielding a shift of electronic momenta analogous to that stemming from a vector potential. As a result, mechanical deformations translate into fictitious gauge fields in the Dirac Hamiltonian. 

In addition, elastic deformations involving local variations of areas induce fluctuations in the electronic density that translate into scalar deformation potentials in each layer.\cite{Mahan,Suzuura} As a result, symmetric deformations between the two layers yield a global scalar potential in the bilayer Hamiltonian. In contrast, antisymmetric distortions yield a difference in the potential between the two layers that mimics the effect of an interlayer electric field. The latter mechanism thus results in a deformation-induced gap in the electronic spectrum even in the absence of external gates.

In monolayer graphene the fictitious vector potential induced by mechanical deformations has been investigated in the past.\cite{Vozmediano,Mahan,Suzuura,Mariani08} The massless Dirac Hamiltonian describing the low energy properties of electrons in a deformed suspended membrane has the form $H^{(+)}_{}=v\mbox{\boldmath$\sigma$}\cdot\left({\bf p}+e{\bf A}^{(+)}_{}\right)$ in the valley $\tau=+$, with $v$ the electron velocity, $\mbox{\boldmath$\sigma$}=(\sigma_{x}^{},\sigma_{y}^{},\sigma_{z}^{})$ the pseudospin vector of Pauli matrices in the sublattice space and ${\bf p}=(p_{x}^{},p_{y}^{},0)$ the electronic momentum. The fictitious vector potential is expressed as ${\bf A}^{(+)}_{}= \sigma_{0}^{}\, \hbar/t\, \left( \partial t/\partial a\right)\left((u_{xx}^{}-u_{yy}^{})/2,-u_{xy}^{}\right)$ in terms of the in-plane hopping energy $t$ and of the strain tensor of the 2D membrane $u_{ij}^{}=(\partial_{i}^{}u_{j}^{}+\partial_{j}^{}u_{i}^{}+\partial_{i}^{}h^{}_{}\, \partial_{j}^{}h^{}_{})/2$, with $\sigma_{0}^{}$ the unit matrix in the sublattice space. Here ${\bf u}({\bf r})$ is the vector field describing in-plane deformations and $h({\bf r})$ the scalar field associated with out-of-plane (flexural) distortions. The symmetry with respect to the plane of graphene is reflected in the quadratic contribution from flexural deformations, as the effects on electrons due to the displacements $h({\bf r})$ and $-h({\bf r})$ are identical.
The Hamiltonian in the other valley ($\tau =-$) is given by $H^{(-)}_{}=H^{(+)*}_{}\big|_{p^{}_{}\rightarrow -p^{}_{}}^{}$, leading to fictitious vector potentials with opposite signs in the two valleys. In particular, the fictitious magnetic fields associated to the vector potentials have opposite signs in the vicinity of the two inequivalent Dirac points, since the effects of deformations do not break time-reversal invariance. 

\subsection{Effects of deformations in the electronic Hamiltonian}

In order to calculate the vector potential in bilayer graphene, we analyse the shift of the atomic positions under a generic distortion. In-plane deformations in layer $l$ are described by the two-dimensional vector field ${\bf u}^{(l)}_{}({\bf r})$, while out-of-plane distortions are associated to a scalar field $h^{(l)}_{}({\bf r})$, so that the atom at position ${\bf r}$ is shifted to ${\bf r}+{\bf u}^{(l)}_{}({\bf r})+\hat{{\bf z}}\, h^{(l)}_{}({\bf r})$. Thus, in the tight-binding Hamiltonian, the hopping term between $|{\bf R}_{sl}^{}\rangle$ and $|{\bf R}_{s'l'}^{}\rangle$ undergoes the change $t_{sl,s'l'}^{}\rightarrow t_{sl,s'l'}^{}+\delta t_{sl,s'l'}^{}$. Here $\delta t_{sl,s'l'}^{}\simeq \left(\partial t_{sl,s'l'}^{}/\partial {\ell}_{sl,s'l'}^{}\right) \delta\ell_{sl,s'l'}^{}$, with 
$\delta\ell_{sl,s'l'}^{}=\big|{\bf R}_{s'l'}^{}-{\bf R}_{sl}^{}+{\bf u}^{(l')}_{}({\bf R}_{s'l'}^{})-{\bf u}^{(l)}_{}({\bf R}_{sl}^{})+\hat{{\bf z}}\, [ h^{(l')}_{}({\bf R}_{s'l'}^{})-h^{(l)}_{}({\bf R}_{sl}^{})]\big|-\big|{\bf R}_{s'l'}^{}-{\bf R}_{sl}^{}\big|$ the variation of the corresponding bond length. 
Among the derivatives $\partial t_{sl,s'l'}^{}/\partial {\ell}_{sl,s'l'}^{}$ only the intralayer one $\partial t_{Al,Bl}^{}/\partial {\ell}_{Al,Bl}^{}\simeq -3\, t_{Al,Bl}^{}/{\ell}_{Al,Bl}^{}$ is known. For all the other cases we will assume typical values $\partial t_{sl,s'l'}^{}/\partial {\ell}_{sl,s'l'}^{}\simeq -\eta_{sl,s'l'}^{} t_{sl,s'l'}^{}/{\ell}_{sl,s'l'}^{}$, with $\eta_{sl,s'l'}^{}$ of order one.
Performing the Bloch band analysis of the deformation-induced corrections to the tight binding problem, we can thus obtain the contribution to the Hamiltonian associated with elastic distortions.
For this purpose it is convenient to introduce symmetric ($S$) and antisymmetric ($A$) deformations in the layer index as follows 
\begin{eqnarray}
&&{\bf u}^{(S/A)}_{}({\bf r})=\frac{{\bf u}^{(2)}_{}({\bf r})\pm {\bf u}^{(1)}_{}({\bf r})}{2}\\
&&h^{(S/A)}_{}({\bf r})=\frac{h^{(2)}_{}({\bf r})\pm h^{(1)}_{}({\bf r})}{2}\; ,\nonumber
\end{eqnarray}
where the upper/lower sign is associated to the symmetric/antisymmetric channel.
In terms of these modes, and using a long-wavelength expansion of the variation of the bond lengths $\delta\ell_{sl,s'l'}^{}$, we deduce the corrections to the Hamiltonian in the two valleys ($\tau =\pm$) in the Dirac matrix formalism as
\begin{equation}
\label{deltaH}
\delta H_{}^{(\tau)}=\left(
\begin{array}{cccc}
D_{1}^{} & F_3^{(\tau)} & F^{(\tau)\dagger}_{4} & F_1^{(\tau)\dagger}\\
F_3^{(\tau)\dagger} & D_{2}^{} & F_2^{(\tau)} & F^{(\tau)}_{4}\\
F^{(\tau)}_{4} & F_2^{(\tau)\dagger} & D_{2}^{} & F_{\gamma}^{(\tau)}\\
F_1^{(\tau)} & F^{(\tau)\dagger}_{4} & F_{\gamma}^{(\tau)} & D_{1}^{}
\end{array}
\right)\; 
\end{equation}
with 
\begin{eqnarray}
\label{Gauge}
&&D^{}_{l =1,2}=g\, {\rm Tr}[u^{(l )}_{ij}] \nonumber\\
&&F^{(\tau)}_{l =1,2}=\frac{3}{4}a\,\frac{\partial t_{l}^{}}{\partial a}\left[ u_{xx}^{(l)}-u_{yy}^{(l)}-i\tau \left(u_{xy}^{(l)}+u_{yx}^{(l)}\right)\right]\nonumber\\
&&F^{(\tau)}_{3}=\frac{3}{2\tilde{c}}\,\frac{\partial \gamma^{}_{3}}{\partial \tilde{c}}\, {\cal F}[{\bf u}^{(S)}_{},\, {\bf u}^{(A)}_{},\, h^{(S)}_{},\, h^{(A)}_{}]\\
&&F^{(\tau)}_{4}=\frac{3}{2\tilde{c}}\,\frac{\partial \gamma_{4}^{}}{\partial \tilde{c}}\, {\cal F}[{\bf u}^{(S)}_{},\, -{\bf u}^{(A)}_{},\, -h^{(S)}_{},\, h^{(A)}_{}]\nonumber\\
&&F^{}_{\gamma}=-2\frac{\partial\gamma}{\partial c}\left[h^{(A)}_{}+\frac{{\bf u}^{(A)2}_{}}{c}\right]\; .\nonumber
\end{eqnarray}
Here $u_{ij}^{(l)}=(\partial_{i}^{}u_{j}^{(l )}+\partial_{j}^{}u_{i}^{(l )}+\partial_{i}^{}h^{(l )}_{}\, \partial_{j}^{}h^{(l )}_{})/2$ is the strain tensor of the two-dimensional membrane in layer $l$ and, in lowest order in the deformation fields, we find 
\begin{eqnarray}
\label{CalF}
&&{\cal F}[{\bf u}^{(S)}_{},\, {\bf u}^{(A)}_{},\, h^{(S)}_{},\, h^{(A)}_{}]= ac\left(\partial_{y}^{}h^{(S)}_{}-i\tau\partial_{x}^{}h^{(S)}_{}\right)\nonumber\\
&&\quad +\frac{a^{2}_{}}{2}\left(u^{(S)}_{xx}-u^{(S)}_{yy}-i\tau\left(u^{(S)}_{xy}+u^{(S)}_{yx}\right)\right)\nonumber\\
&&\quad +2a\left(u^{(A)}_{y}-i\tau u_{x}^{(A)}\right)\; .
\end{eqnarray}
Thus the correction terms in the two valleys $\tau =\pm$ are related by the symmetry $\delta H^{(-)}_{}=\delta H^{(+)*}_{}$.
In Eq.\ (\ref{deltaH}) we also introduced the term $D^{}_{l}$ representing the deformation potential for the layer $l$ associated to local variation of areas in a distorted elastic medium, \cite{Suzuura} with $g$ the deformation potential coupling constant. While the bare value of $g$ is estimated around $20-30\, {\rm eV}$,\cite{Suzuura} in graphene samples the coupling constant is reduced by screening so that it effectively depends on the electron density.\cite{Mariani10,Hwang07} In contrast, gauge fields are not affected by screening.\cite{FelixPacoEros} As a result, one expects that the vector potential dominates in (significantly strongly) doped graphene, while the deformation potential would be dominant in the immediate vicinity of the Dirac point. 

\begin{figure}[ht]
	\centering
		\includegraphics[width=1.0\columnwidth]{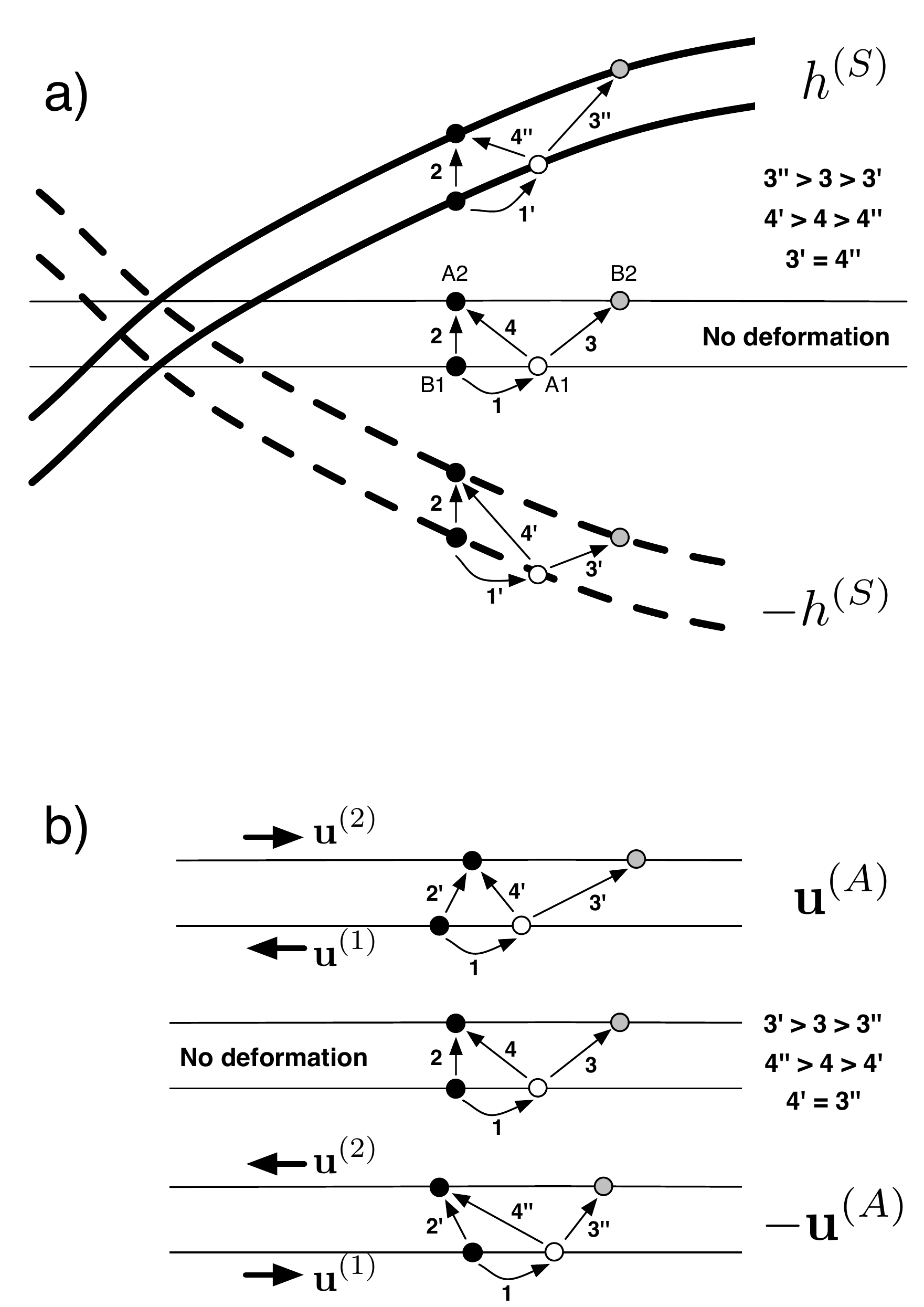}
			\caption{Schematic description of the effect of deformations on bond lengths. This side view, taken along the direction ${\bf e}_{j}^{}$, shows the atoms involved in intra and interlayer hopping processes in a unit cell. a) Bilayer under flexural deformations $h^{(S)}_{}$ and $-h^{(S)}_{}$. The difference between the hopping lengths ${\bf 3'}$ and ${\bf 3''}$ (as well as between ${\bf 4'}$ and ${\bf 4''}$) breaks the symmetry with respect to the plane and is responsible for the appearance of a linear coupling with $h^{(S)}_{}$ in the gauge field $F_{3}^{(\tau)}$ (and $F_{4}^{(\tau)}$). The fact that ${\bf 3'}={\bf 4''}$ leads to the symmetry $h^{(S)}_{}\rightarrow -h^{(S)}_{}$ between $F_{3}^{(\tau)}$ and $F_{4}^{(\tau)}$. b) same as in a), but for in-plane antisymmetric deformations ${\bf u}^{(A)}_{}$ and $-{\bf u}^{(A)}_{}$. The equality ${\bf 4'}={\bf 3''}$ accounts for the symmetry ${\bf u}^{(A)}_{}\rightarrow -{\bf u}^{(A)}_{}$ between $F_{3}^{(\tau)}$ and $F_{4}^{(\tau)}$. \label{SymmetryNew}}
\end{figure}

The terms $F^{(\tau)}_{l =1,2}$ in Eq.\ (\ref{Gauge}) are the same as those appearing in monolayer graphene. They are linear in the in-plane deformations causing variations in the bond-lengths. However, the symmetry with respect to the $x-y$ plane forces the coupling with out-of-plane deformations to be quadratic.
In contrast, in the terms $F^{(\tau)}_{3}$ and $F^{(\tau)}_{4}$ it is interesting to notice the appearance of a linear coupling between electrons and symmetric flexural deformations ($h_{}^{(S)}$) for the skewed interlayer hopping terms ($A1-A2,\, B1-B2$ and $A1-B2$).
In view of the results for the monolayer, protected by the symmetry with respect to the plane, it may be surprising that such a term exists. However, due to the rotation between the layers involved in the Bernal stacking, the two layers are not equivalent. Thus, for skewed interlayer bonds, the symmetric flexural deformations $h^{(S)}_{}$ and $-h^{(S)}_{}$ induce different effects, resulting in the appearance of a linear residual contribution in the fictitious gauge fields. In more detail, this can be understood by considering the bond lengths involved in the interlayer hopping terms different from the ``vertical" $B1-A2$ one. In Fig.\ \ref{SymmetryNew} the interlayer skewed bonds are illustrated under a generic $h^{(S)}_{}$ (and $-h^{(S)}_{}$) deformation. The two are evidently different, due to the shift between the projected positions of the atoms involved. 
The analysis of the figure also reveals why the $F^{(\tau)}_{3}$ and $F^{(\tau)}_{4}$ terms differ by the replacements $h^{(S)}_{}\rightarrow -h^{(S)}_{}$ and ${\bf u}^{(A)}_{}\rightarrow -{\bf u}^{(A)}_{}$, as these preserve the bond lengths involved in the corresponding hopping terms.

As far as long wavelength antisymmetric flexural deformations $h_{}^{(A)}$ are concerned, they correspond to a local modulation of the interlayer distance which preserves the structure of Bernal stacking. Their only effect is thus to induce a numerical renormalisation of the velocities $v_{3}^{}$ and $v_{4}^{}$. In Eqs.\ (\ref{Gauge}) and (\ref{CalF}) this would merely lead to a sub-leading correction in $h^{(A)}_{}/\tilde{c}\ll 1$ and $|p|a/\hbar\ll 1$ of the form $\delta {\cal F}=(2ac/\hbar)\, h^{(A)}_{}\left(\tau p_{x}^{}+ip_{y}^{}\right)$, which can be neglected.

Finally, the term $F_{\gamma}^{}$ is associated to the vertical interlayer hopping $B1-A2$, which does not involve any skewed bond. As a consequence it is not affected by symmetric deformations $h_{}^{(S)}$ and ${\bf u}^{(S)}_{}$. The variation of the bond length stems uniquely from antisymmetric deformations. It is linear in the flexural distortions $h^{(A)}_{}$ and quadratic in the in-plane ones ${\bf u}^{(A)}_{}$.

The energy corrections $F_{j}^{(\tau)}$ (with $j=1,...,4$) in $\delta H_{}^{(\tau)}$ thus affect the unperturbed terms $v_{j}^{}p$ in $H_{0}^{(\tau)}$ as fictitious gauge fields acting on the electronic orbital degrees of freedom. The symmetry $H_{0}^{(-)}+\delta H_{}^{(-)}=H_{0}^{(+)*}\big|_{p^{}_{}\rightarrow -p^{}_{}}^{}+\delta H_{}^{(+)*}$, together with $F_{j}^{(-)}=F_{j}^{(+)*}$ reveals that the fictitious gauge fields have opposite signs in the two valleys, as in the monolayer case. As a consequence, the fictitious magnetic fields generated by elastic deformations are also opposite in the two valleys, as requested by the fact that elastic deformations do not break time-reversal invariance.

\subsection{Explicit form of the fictitious vector potential}

In order to express the effect of deformations on the electronic momenta in terms of a vector potential, we can rewrite the total Hamiltonian $H_{0}^{(+)}+\delta H_{}^{(+)}$ in the language of Pauli matrices acting on the layer space ($\Sigma_{\alpha}^{}$, with $\alpha \in \{0,x,y,z\}$) and on the sublattice space ($\sigma_{\alpha}^{}$) as 
\begin{eqnarray}
&&H_{0}^{(+)}+\delta H_{}^{(+)}=H_{p}^{(+)}+H_{\gamma}^{}+H_{D}^{}\nonumber \\
&&H_{p}^{(+)}=H_{0,p}^{(+)}+\delta H_{p}^{(+)} \\
&&H_{\gamma}^{}=-\frac{\gamma -F_{\gamma}^{}}{2}\left(\Sigma_{x}^{}\otimes \sigma_{x}^{}+\Sigma_{y}^{}\otimes \sigma_{y}^{}\right)\nonumber \\
&&H_{D}^{}=D^{(S)}_{}\Sigma_{0}^{}\otimes \sigma_{0}^{}-D^{(A)}_{}\Sigma_{z}^{}\otimes \sigma_{0}^{}\; ,\nonumber
\end{eqnarray}
with $\Sigma_{0}^{}$ and $\sigma_{0}^{}$ the identity matrices in the corresponding spaces.
The term $H_{0,p}^{(+)}$ collects all contributions linear in $v_{j}^{}p$ and $\delta H_{p}^{(+)}$ the corrections $F_{j}^{}$ (with $j=1,...,4$), while $H_{\gamma}^{}$ collects the terms involving $\gamma$ and $F_{\gamma}^{}$, and $H_{D}^{}$ those related to the deformation potentials.
In the latter we introduced the symmetric and antisymmetric components $D^{(S)}_{}=(D_{1}^{}+D_{2}^{})/2$ and $D^{(A)}_{}=(D_{2}^{}-D_{1}^{})/2$, corresponding to different variations of areas in the two layers. In analogy with the monolayer case, the Hamiltonian $H_{p}^{(+)}$ can be written as 
\begin{equation}
H_{p}^{(+)}={\bf V}^{(+)}_{}\cdot ({\bf p}+e{\bf A}^{(+)}_{})
\end{equation}
with $H_{0,p}^{(+)}\equiv {\bf V}^{(+)}_{}\cdot {\bf p}$ expressed in terms of the vector of velocity matrices
\begin{eqnarray}
&&{\bf V}^{(+)}_{}=(V_{x}^{(+)},V_{y}^{(+)})\; , \quad {\rm with}\\
&&V_{x}^{(+)}=\frac{v_{1}^{}+v_{2}^{}}{2}\,\Sigma_{0}^{}\otimes\sigma_{x}^{}+\frac{v_{1}^{}-v_{2}^{}}{2}\,\Sigma_{z}^{}\otimes\sigma_{x}^{}\nonumber\\
&&\quad\quad\quad +\frac{v_{3}^{}}{2}\left(\Sigma_{x}^{}\otimes\sigma_{x}^{}-\Sigma_{y}^{}\otimes\sigma_{y}^{}\right)+v_{4}^{}\Sigma_{x}^{}\otimes\sigma_{0}^{}\nonumber \\
&&V_{y}^{(+)}=\frac{v_{1}^{}+v_{2}^{}}{2}\,\Sigma_{0}^{}\otimes\sigma_{y}^{}+\frac{v_{1}^{}-v_{2}^{}}{2}\,\Sigma_{z}^{}\otimes\sigma_{y}^{}\nonumber\\
&&\quad\quad\quad -\frac{v_{3}^{}}{2}\left(\Sigma_{x}^{}\otimes\sigma_{y}^{}+\Sigma_{y}^{}\otimes\sigma_{x}^{}\right)+v_{4}^{}\Sigma_{y}^{}\otimes\sigma_{0}^{}\; .\nonumber 
\end{eqnarray}
In parallel, for the term $\delta H_{p}^{(+)}$, by direct inspection one finds
\begin{eqnarray}
&&\delta H_{p}^{(+)}\equiv e{\bf V}^{(+)}_{}\cdot {\bf A}^{(+)}_{}=\delta H_{{\rm Re},p}^{(+)}+\delta H_{{\rm Im},p}^{(+)}\;, \quad {\rm with}\\
&&\delta H_{{\rm Re},p}^{(+)}={\rm Re}[F_{1}^{(+)}]\,\frac{\Sigma_{0}^{}+\Sigma_{z}^{}}{2}\otimes\sigma_{x}^{}+{\rm Re}[F_{2}^{(+)}]\,\frac{\Sigma_{0}^{}-\Sigma_{z}^{}}{2}\otimes\sigma_{x}^{}\nonumber\\
&&\quad\quad\quad +{\rm Re}[F_{3}^{(+)}]\,\frac{\Sigma_{x}^{}\otimes\sigma_{x}^{}-\Sigma_{y}^{}\otimes\sigma_{y}^{}}{2}+{\rm Re}[F_{4}^{(+)}]\,\Sigma_{x}^{}\otimes\sigma_{0}^{}\nonumber\\
&&\delta H_{{\rm Im},p}^{(+)}={\rm Im}[F_{1}^{(+)}]\,\frac{\Sigma_{0}^{}+\Sigma_{z}^{}}{2}\otimes\sigma_{y}^{}+{\rm Im}[F_{2}^{(+)}]\,\frac{\Sigma_{0}^{}-\Sigma_{z}^{}}{2}\otimes\sigma_{y}^{}\nonumber\\
&&\quad\quad\quad -{\rm Im}[F_{3}^{(+)}]\,\frac{\Sigma_{x}^{}\otimes\sigma_{y}^{}+\Sigma_{y}^{}\otimes\sigma_{x}^{}}{2}+{\rm Im}[F_{4}^{(+)}]\,\Sigma_{y}^{}\otimes\sigma_{0}^{}\nonumber\; ,
\end{eqnarray}
leading to the vector potential
\begin{eqnarray}
&&e{\bf A}^{(+)}_{}=\left(eA^{(+)}_{x},eA^{(+)}_{y}\right)\; ,\\
&&eA^{(+)}_{x}=\left(V_{x}^{(+)} \right)^{-1}_{} \delta H_{{\rm Re},p}^{(+)}\nonumber \\
&&eA^{(+)}_{y}=\left(V_{y}^{(+)} \right)^{-1}_{} \delta H_{{\rm Im},p}^{(+)}\; .\nonumber
\end{eqnarray}
In contrast to monolayers, for bilayer graphene in the general case (with $v_{1}^{}\neq v_{2}^{}$, $v_{3}^{}\neq 0$, $v_{4}^{}\neq 0$ and all the associated corrections $F_{j}^{(\tau)}$) the analytical expression for the vector potential is quite cumbersome, due to the matrix structure of the velocity vector ${\bf V^{(+)}_{}}$. For illustration purposes we consider the simpler case $v_{1}^{}=v^{}_{2}=v$, $F^{(+)}_{1}=F^{(+)}_{2}=F^{(+)}_{}$ and $v_{4}^{}=F^{(+)}_{4}=0$ leading to the compact expression
\begin{eqnarray}
&&eA^{(+)}_{x}={\rm Re}[F^{(+)}_{}]\left(\frac{1}{v}\, \Sigma_{0}^{}\otimes\sigma_{0}^{}-\frac{v_{3}^{}}{2v^{2}_{}}\left(\Sigma_{x}^{}\otimes\sigma_{0}^{}-i\Sigma_{y}^{}\otimes\sigma_{z}^{}\right)\right)\nonumber \\
&&\quad\quad+{\rm Re}[F^{(+)}_{3}]\left(\frac{1}{2v}\left(\Sigma_{x}^{}\otimes\sigma_{0}^{}-i\Sigma_{y}^{}\otimes\sigma_{z}^{}\right)\right) \\
&&eA^{(+)}_{y}={\rm Im}[F^{(+)}_{}]\left(\frac{1}{v}\, \Sigma_{0}^{}\otimes\sigma_{0}^{}+\frac{v_{3}^{}}{2v^{2}_{}}\left(\Sigma_{x}^{}\otimes\sigma_{0}^{}-i\Sigma_{y}^{}\otimes\sigma_{z}^{}\right)\right)\nonumber \\
&&\quad\quad-{\rm Im}[F^{(+)}_{3}]\left(\frac{1}{2v}\left(\Sigma_{x}^{}\otimes\sigma_{0}^{}-i\Sigma_{y}^{}\otimes\sigma_{z}^{}\right)\right)\nonumber\; .
\end{eqnarray}
The terms proportional to $v_{3}^{}$ and $F^{(+)}_{3}$ are related to interlayer hopping processes and are associated with the appearance of the Pauli matrices $\Sigma_{x,y}^{}$ which involve mixing of the two layers. In contrast, the special case $v_{3}^{}=F^{(+)}_{3}=0$ would lead to two decoupled layers as far as $H_{p}^{(+)}$ is concerned, each characterised by the vector potential $e{\bf A}^{(+)}_{}=\sigma_{0}^{}/v\, \left({\rm Re}[F^{(+)}_{}],\, {\rm Im}[F^{(+)}_{}]\right)$. Using the expression for $F^{(+)}_{1,2}$ in Eq.\ (\ref{Gauge}) and the velocity $v=3at/2\hbar$ this indeed coincides with the vector potential of a monolayer. The coupling between the layers would still be present via the term $H_{\gamma}^{}$ which, however, does not involve a vector potential.

\subsection{Effective low-energy Hamiltonian}

The $4\times 4$ Hamiltonians $H_{0}^{(\tau)}+\delta H_{}^{(\tau)}$ in Eqs.\ (\ref{H0+}) and (\ref{deltaH}) contain the complete information concerning the properties of electrons and of their coupling to elastic deformations. The unperturbed Hamiltonian $H_{0}^{(\tau)}$ is diagonalised in terms of four electronic bands, two of them touching at low energy, and two describing split modes at high-energies of order $\pm \gamma$, due to the ``vertical" $B1-A2$ interlayer hopping.

In order to analyse the low-energy sector of the spectrum, we can produce an effective $2\times 2$ Hamiltonian in the $A1-B2$ subspace along the same line as done, e.g. in Refs.\ \onlinecite{McCann,SchriefferWolfNote}. We consider the original $4\times 4$ Hamiltonian $H^{(\tau )}_{}=H^{(\tau )}_{0}+\delta H^{(\tau )}_{}$ as made of four $2\times 2$ blocks $H^{(\tau )}_{ij}$, ($i, j\in {1,2}$) with the upper left block $H_{11}^{(\tau )}$ describing the low energy sector. We introduce the matrix Green's function $G_{}^{(\tau )}=\left(\epsilon {\bf 1}-H^{(\tau )}_{}\right)^{-1}_{}$, with ${\bf 1}$ the unit matrix, and by direct matrix inversion we get $G_{11}^{(\tau )}=\left(\epsilon {\bf 1}-H_{11}^{(\tau )}-H_{12}^{(\tau )}(\epsilon {\bf 1}-H_{22}^{(\tau )})^{-1}_{}H_{21}^{(\tau )}\right)^{-1}_{}$. Thus the effective low energy Hamiltonian is obtained as $H_{{\rm eff}}^{(\tau )}=\epsilon {\bf 1}-G_{11}^{(\tau )-1}\simeq H_{11}^{(\tau )}-H_{12}^{(\tau )}(H_{22}^{(\tau )})^{-1}_{}H_{21}^{(\tau )}$ in the limit $\epsilon\ll \gamma$. By keeping only the lowest non-vanishing order in $F^{(\tau )}_{j}/\gamma \ll 1$ and $D^{}_{l}/\gamma \ll 1$, after lengthy but straightforward steps, we obtain the effective low-energy Hamiltonian in the "+" valley
\begin{widetext}
\begin{equation}
\label{Hlowdef}
H_{{\rm eff}}^{(+)}=\left(
\begin{array}{cc}
D^{(S)}_{}-D^{(A)}_{}+\frac{\Delta}{2} & v_3^{}P_{3}^{(+)}\\
v_3^{}P_{3}^{(+)\dagger} & D^{(S)}_{}+D^{(A)}_{}-\frac{\Delta}{2}
\end{array}
\right)\; + \frac{1}{\gamma}
\left(
\begin{array}{cc}
v_{1}^{}v_{4}^{}\left(P_{4}^{(+)\dagger}P_{1}^{(+)}+P_{1}^{(+)\dagger}P_{4}^{(+)}\right) & v^{2}_{4}\left(P_{4}^{(+)\dagger}\right)^{2}_{}+v_{1}^{}v_{2}^{}P_{1}^{(+)\dagger}P_{2}^{(+)\dagger}\\
v^{2}_{4}\left(P_{4}^{(+)}\right)^{2}_{}+v_{1}^{}v_{2}^{}P_{2}^{(+)}P_{1}^{(+)} & v_{2}^{}v_{4}^{}\left(P^{(+)}_{2}P_{4}^{(+)\dagger}+P_{4}^{(+)}P_{2}^{(+)\dagger}\right)
\end{array}
\right)
\end{equation}
\end{widetext}
where we introduced the kinetic momenta $P_{j}^{(\tau)}=p+F_{j}^{(\tau)}/v_{j}^{}$ for $j=1,...,4$. The Hamiltonian in the "$-$" valley is then expressed as  $H_{{\rm eff}}^{(-)}=H_{{\rm eff}}^{(+)*}\big|_{p^{}_{}\rightarrow -p^{}_{}}^{}$. The term $F_{\gamma}^{}$ yields a small correction to the hopping energy $\gamma$ that affects the high energy bands. However, it produces only sub-leading corrections of order $F^{}_{\gamma}/\gamma\ll 1$ to the low energy Hamiltonian (\ref{Hlowdef}), which are thus neglected.

In the effective Hamiltonian (\ref{Hlowdef}) we also included an on-site energy difference $\Delta$ between the two layers describing the effect of an interlayer electric field. This term, together with the antisymmetric component of the deformation potential $D^{(A)}_{}$ has the physical effect of inducing a gap in the electronic spectrum. Thus, our results show that such a gap is in principle realisable, without any inter-layer electric field, under different variation of local areas for the two layers.

The effective Hamiltonian (\ref{Hlowdef}) is the main result of this paper. It can be used as the starting point for the investigation of electromechanical properties of bilayer graphene under arbitrary elastic deformations. Without loss of generality, from now on we focus on one valley ($\tau = +$).
We first review the electronic spectrum in the absence of mechanical deformations, and then briefly discuss the effect of strain configurations inducing uniform fictitious gauge fields. The corresponding effects on the low-energy electronic band-structure of bilayer graphene have been recently discussed by other authors as well.\cite{Korean,Falko11}  As a further application we will then use the Hamiltonian (\ref{Hlowdef}) to describe the electron-phonon coupling and its contribution to the resistivity of suspended bilayer graphene.

\section{Band structure without deformations}
\label{sec:NoStrain}

In the absence of mechanical deformations the effective Hamiltonian (\ref{Hlowdef}) reduces to 
\begin{eqnarray}
\label{Hlow}
H_{{\rm eff}}^{(+)}&=&\left(
\begin{array}{cc}
\Delta/2 & v_3^{}p\\
v_3^{}p^{\dagger}_{} & -\Delta/2
\end{array}
\right)\\
&+&\frac{1}{\gamma}
\left(
\begin{array}{cc}
2v_{1}^{}v_{4}^{}\left|p\right|^{2}_{} & (v_{4}^{2}+v_{1}^{}v_{2}^{})p^{\dagger 2}_{}\\
(v_{4}^{2}+v_{1}^{}v_{2}^{})p^{2}_{} & 2v_{2}^{}v_{4}^{}\left|p\right|^{2}_{}
\end{array}
\right)\; .\nonumber
\end{eqnarray}
Different cases can be analysed, according to the importance of the various interlayer hopping terms and to the asymmetry in the intralayer velocities $v_{1}^{}$ and $v_{2}^{}$. Before discussing the most general case, we briefly review the features stemming from the successive introduction of the hopping terms, in order of descending magnitude.

i) Most results on bilayer graphene at energies not too close to the Dirac point can be analysed by considering only the dominant $A2-B1$ interlayer hopping term $\gamma$, with equal intralayer velocities $v_{1}^{}=v_{2}^{}=v\simeq 8\cdot 10^{5}_{}\, {\rm m\cdot s^{-1}_{}}$, thus neglecting $v_{3}^{}$ as well as $v_{4}^{}$ in Eq.\ (\ref{Hlow}). In the absence of interlayer electric fields ($\Delta =0$) this yields a parabolic spectrum with two energy bands $\epsilon_{\pm}^{}=\pm v^{2}_{}|p|^{2}_{}/\gamma\equiv \pm |p|^{2}_{}/2m$ characterised by an effective mass $m=\gamma /2v^{2}_{}\simeq 0.05\, m_{{\rm e}}^{}$ ($m_{{\rm e}}^{}$ the free electron mass).\cite{McCann,Nilsson} The two bands touch at one Dirac point at $|p|=0$. 
Using the complex representation of the momentum $p=\left| p\right| \exp[i\phi]$ the effective Hamiltonian can be written as 
\begin{equation}
H_{{\rm eff}}^{(+)}\simeq\frac{|p|^{2}_{}}{2m}\, \mbox{\boldmath$\sigma$}\cdot (\cos 2\phi ,\sin 2\phi, 0)\; ,\nonumber
\end{equation}
with $\mbox{\boldmath$\sigma$}=(\sigma_{x}^{},\sigma_{y}^{},\sigma_{z}^{})$ the pseudospin vector of Pauli matrices in the sublattice space.
Comparing this to the electronic energy, we obtain the chirality condition $\mbox{\boldmath$\sigma$}\cdot (\cos 2\phi ,\sin 2\phi, 0)=\pm 1$, with the upper/lower sign corresponding to the conduction/valence band. Thus electrons behave as pseudospin-$1/2$ massive chiral particles, their pseudospin winding by $4\pi$ anticlockwise when $\phi$ goes from $0$ to $2\pi$. This corresponds to a Berry phase of $\Phi =s\, \Omega =2\pi$,\cite{Berry,Carlo} where $s=1/2$ is the particle pseudospin and $\Omega = 4\pi$ the solid angle enclosed by the pseudospin vector while the electronic state is transported anticlockwise in a closed loop in the 2D momentum space around the Dirac point. 

\begin{figure}[ht]
	\centering
		\includegraphics[width=1.0\columnwidth]{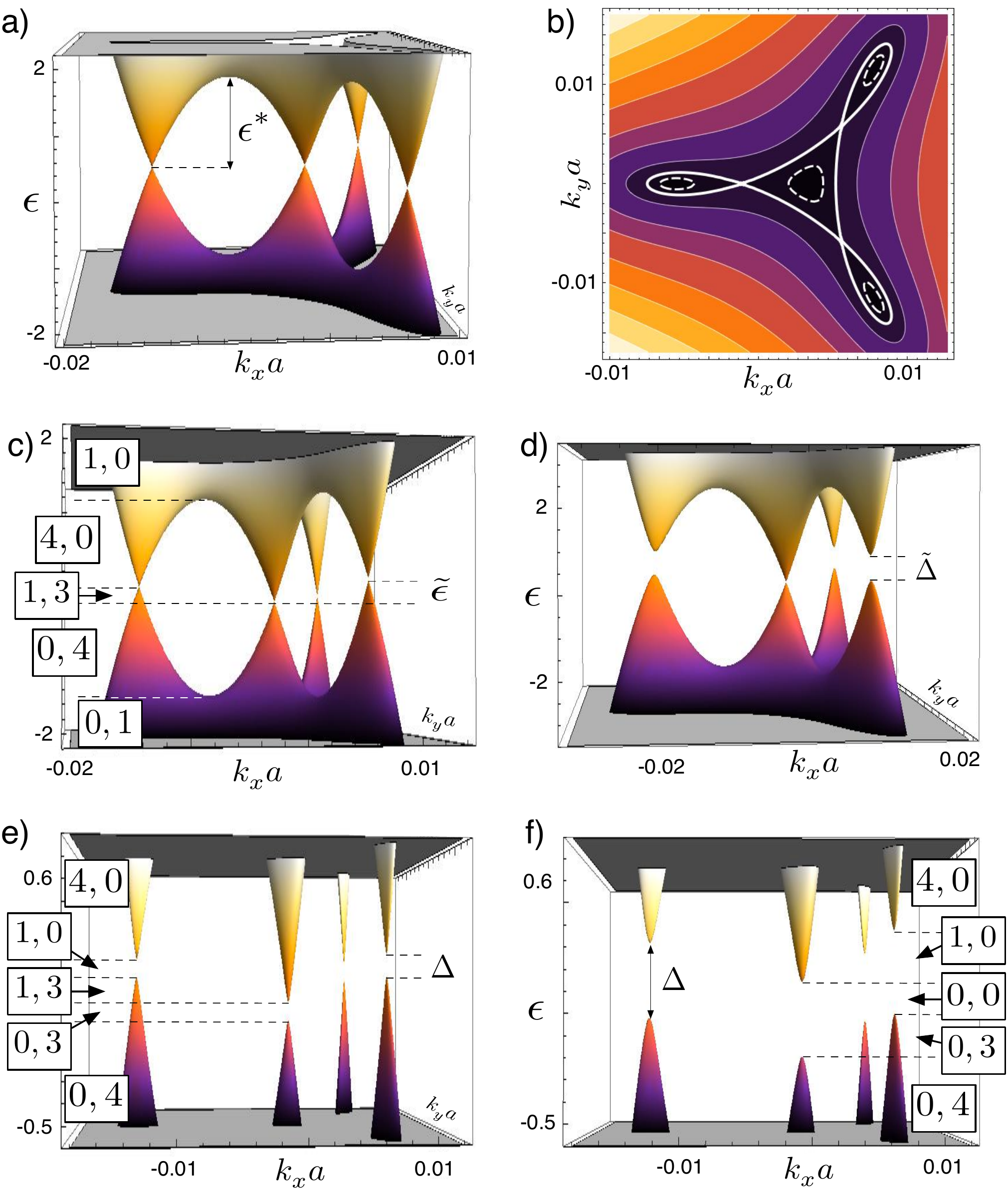}
			\caption{Electronic band structure of bilayer graphene without deformations. All energy scales $\epsilon$ are in ${\rm meV}$. a) The low-energy spectrum neglecting terms in $v_{4}^{}$. Four massless cones touch at zero energy. b) Equipotential lines for the panel a). Dashed lines correspond to $\epsilon <\epsilon^{*}_{}$, yielding four disconnected electron pockets. The thick line corresponds to $\epsilon =\epsilon^{*}_{}$ where the LT occurs. All other continuous lines are at $\epsilon >\epsilon^{*}_{}$, yielding a single connected electron pocket. Dark areas correspond to states close to zero energy while light ones are for higher energies. c) Band structure including $v_{4}^{}$. The central Dirac cones touch at zero energy, while the other three touch at $\tilde{\epsilon}$. The boxes $[N_{e}^{},N_{h}^{}]$ in different energy windows indicate that the FS is made out of $N_{e}^{}$ electron pockets and $N_{h}^{}$ hole pockets. d) Same as c), but with asymmetric intralayer velocities, corresponding to $t_{2}^{}=t_{1}^{}/4=1\, {\rm eV}$. This large asymmetry is used to stress the formation of the minigaps $\tilde{\Delta}$. e) Magnification of the low energy spectrum in c), but with a small interlayer gap $\Delta < \tilde{\epsilon}$. f) Same as in e), but with a larger interlayer gap $\Delta > \tilde{\epsilon}$.	\label{NoStrain}}
\end{figure}

In the presence of a perpendicular magnetic field this spectrum is characterised by a doubly-degenerate Landau-level at zero energy (per spin and per valley).\cite{Geim,McCann} Finally, an interlayer electric field ($\Delta \neq 0$) yields an energy gap $\Delta$ in the spectrum.\cite{McCann,Nilsson}

ii) The next step in the hierarchy of approximations is to include the terms proportional to $v_{3}^{}\simeq 10^{5}_{}\, {\rm m}\cdot {\rm s}^{-1}_{}$ in Eq.\ (\ref{Hlow}), while still neglecting $v_{4}^{}$ and keeping $v_{1}^{}=v_{2}^{}=v$.
In this case, and for $\Delta =0$, one obtains two energy bands with dispersion
\begin{equation}
\epsilon_{\pm}^{}=\pm \left|v_{3}^{}p^{\dagger}_{}+\frac{p^{2}_{}}{2m}\right|\; .\nonumber
\end{equation}
The two bands touch at zero energy at four Dirac points, obtained by imposing $\epsilon_{\pm}^{}=0$. The four points are given by $|p|=0$ as well as $|p|=2mv_{3}^{}$ and $\phi =\pi (2n+1)/3$, with $n=0,1,2$, highlighting the latent symmetry of the honeycomb lattice.
Around each Dirac point the dispersion is massless. It is isotropic around the central point at $|p|=0$ and anisotropic for the three satellite cones, as illustrated in Figs.\ \ref{NoStrain}a and \ref{NoStrain}b.\cite{McCann,Nilsson} The linearisation of the Hamiltonian near the new Dirac points will be presented at the next level of approximation, where $v_{4}^{}$ is also included. 
\\
The four cones meet at energy $\epsilon^{*}_{}= \gamma v^{2}_{3}/4v^{2}_{}=mv^{2}_{3}/2\simeq 1.6\, {\rm meV}$ so that at higher energies the spectrum is essentially parabolic. 

As function of the electron density (i.e.\ of the Fermi energy $\epsilon_{{\rm F}}^{}$), the Fermi sea (FS) changes its shape and topology. 
In each valley, we can compactly denote the typology of a FS characterised by $N_{e}^{}$ electron pockets and $N_{h}^{}$ hole pockets by the notation $[N_{e}^{},N_{h}^{}]$. 
For $|\epsilon_{{\rm F}}^{}|> \epsilon_{}^{*}$, corresponding to a concentration of electrons (or holes) larger than $(2/\pi^2)(mv_{3}^{}/\hbar)^{2}_{}\simeq 2\cdot 10^{10}_{}\, {\rm cm}^{-2}_{}$, the FS in each valley is connected and topologically equivalent to a circle (i.e. of type $[1,0]$ or $[0,1]$). At $|\epsilon_{{\rm F}}^{}|= \epsilon_{}^{*}$ a Lifshitz transition (LT) occurs and the FS shows knots that develop into four disconnected electron (hole) pockets for $0<|\epsilon_{{\rm F}}^{}|<\epsilon_{}^{*}$ (i.e. type $[4,0]$ or $[0,4]$). The single-particle electronic density of states (DOS) vanishes linearly while approaching zero doping and exhibits a peak at the LT, that should lead to observable features, e.g. in compressibility measurements or the transport properties as a function of the carrier density. The topology of the Fermi surface close to the Lifshitz transition is presented in Fig.\ \ref{NoStrain}b.

In the presence of a perpendicular magnetic field, each massless Dirac cone yields one Landau-level at zero energy, as in monolayer graphene, leading to a four-fold degeneracy per valley and spin. This scenario should result in quantum Hall plateaux in the transverse conductivity at $\pm 8\, e^{2}_{}/h$ for small enough magnetic fields and close to zero carrier density.\cite{McCann} 

It has to be pointed out that $\epsilon^{*}_{}$ is a rather small energy scale. The physics of the LT can thus be observed only in extremely clean samples at low density, so that the smearing due to disorder does not obscure the pertinent features. An alternative possibility to overcome this difficulty would be to tune the LT to higher energies. This could be achieved in bilayer samples by inducing significant strain (see Sec.\ \ref{sec:Strain}). Alternatively, ABC-stacked trilayer graphene shows a LT around $10\, {\rm meV}$.\cite{McCannTrilayer} The larger energy associated with this LT allows for its experimental observation in samples with high mobility, as reported recently.\cite{LauABC} However, as the LT in trilayers originates from the vertical hopping between the first and the third layer, it should be only weakly sensitive to deformations and it could not be easily tuned as in bilayers.

iii) If we still assume $v_{1}^{}=v_{2}^{}=v$, but we do not neglect the $A1-A2$ and $B1-B2$ interlayer hoppings, taking into account $v_{4}^{}\simeq 1.4\cdot 10^{4}_{}\, {\rm m}\cdot {\rm s}^{-1}_{}$ in Eq.\ (\ref{Hlow}), we still obtain four massless Dirac cones at low energy, but the two bands \emph{do not} touch at the same energy.\cite{Mikitik}
Indeed, for $\Delta =0$, they are given by
\begin{equation}
\epsilon_{\pm}^{}=\frac{2v_{4}^{}}{v}\frac{\left| p\right|^{2}_{}}{2m}\pm \left|v_{3}^{}p^{\dagger}_{}+\frac{p^{2}_{}}{2m}\left(1+\frac{v_{4}^{2}}{v^{2}_{}}\right)\right|\; .\nonumber
\end{equation}
The four Dirac points are found at $|p|=0$, where the bands touch at zero energy, as well as $|p|=2mv_{3}^{} /(1+v_{4}^{2}/v^{2}_{})$ with $\phi =\pi (2n+1)/3$, where the bands touch at energy $\tilde{\epsilon}=2\gamma vv_{3}^{2}v_{4}^{}/(v^{2}_{}+v_{4}^{2})^{2}_{}=4mv_{3}^{2}v_{4}^{}/\left[v\left(1+v_{4}^{2}/v^{2}_{}\right)^2_{}\right]$. Since $v_{4}^{}/v\ll 1$ we have $\tilde{\epsilon}\simeq 8\epsilon^{*}_{}\, v_{4}^{}/v\simeq 0.2\, {\rm meV}$. This dispersion is illustrated in Fig.\ \ref{NoStrain}c.

We can explicitly expand the Hamiltonian around the different Dirac points (described by the complex momenta $p_{{\rm D}}^{}$) by considering $p=p_{{\rm D}}^{}+\delta p$.
Around $p_{{\rm D}}^{}=0$ the linearised Hamiltonian is given by
\begin{equation}
H_{{\rm eff}}^{(+)}\simeq\left(
\begin{array}{cc}
0 & v_3^{}\delta p\\
v_3^{}\delta p^{\dagger}_{} & 0
\end{array}
\right)=v_{3}^{}\left|\delta p\right|\, \mbox{\boldmath$\sigma$}\cdot (\cos \phi ,-\sin \phi, 0)\nonumber
\end{equation}
and describes massless chiral fermions with Berry phase $-\pi$, due to the clockwise winding of the spinor for an anticlockwise loop of $\delta p$.
In a similar way, the expansion around one of the other Dirac points, e.g. $p_{{\rm D}}^{}=-2mv_{3}^{} /(1+v_{4}^{2}/v^{2}_{})$, yields
\begin{eqnarray}
&&H_{{\rm eff}}^{(+)}\simeq \tilde{\epsilon}{\bf 1}-v_{3}^{}\left(
\begin{array}{cc}
0 & \delta p_{x}^{}-3i\delta p_{y}^{}\nonumber \\
\delta p_{x}^{}+3i\delta p_{y}^{} & 0
\end{array}
\right)\\
&&\quad =\tilde{\epsilon}{\bf 1}-v_{3}^{}\left|\delta p\right|\, \mbox{\boldmath$\sigma$}\cdot (\cos \phi ,3\sin \phi, 0)\; ,\nonumber
\end{eqnarray}
describing massless Dirac fermions with Berry phase $\pi$ and elliptical equipotential lines (see Fig.\ \ref{NoStrain}b).

As a function of the carrier density, the FS develops interesting structures. Due to the fact that the massless Dirac cones touch at different energies, the DOS never vanishes. For $0<\epsilon_{{\rm F}}^{}<\tilde{\epsilon}$ the FS is of type $[1,3]$. On the other hand, the spectrum at energies larger than $\tilde{\epsilon}$ and the LT remain essentially unaffected (see Fig.\ \ref{NoStrain}c). The critical energy for the occurrence of the LT is slightly renormalised to $\epsilon^{*}_{+}\simeq\epsilon^{*}_{}\left(1+2v_{4}^{}/v\right)$ for electron doping and to $\epsilon^{*}_{-}\simeq -\epsilon^{*}_{}\left(1-2v_{4}^{}/v\right)$ for hole doping.

In the presence of an external magnetic field, Landau-levels corresponding to massless Dirac fermions are generated in the low energy sector. However, due to the energy offset $\tilde{\epsilon}$, there is only one Landau level at zero energy (per valley and per spin) stemming from the central Dirac cone. 
This mechanism would result in plateaux of the Hall conductivity at $\pm e^{2}_{}/h$ close to zero doping at low magnetic fields in extremely high mobility samples.  
However, due to the smallness of the energy scale $\tilde{\epsilon}$, the shift of the three satellite Dirac cones with respect to the central one is not observable with the present quality of bilayer samples. 

It is interesting to notice that a finite value of interlayer electric field corresponding to $0<\Delta\leq \tilde{\epsilon}$ does not induce a global gap in the spectrum, but rather opens a gap in each individual Dirac cone. A global gap opens up only for $\Delta> \tilde{\epsilon}$.
Thus, in the presence of $\Delta\neq 0$, one can identify seven regions where the structure of the FS is as follows: \\
1) $[1,0]$ for $\epsilon_{{\rm F}}^{} > \tilde{\epsilon}/4+\sqrt{(\Delta/2)^{2}_{}+(\epsilon^{*}_{+})^{2}_{}}$, \\
2) $[4,0]$ for $\epsilon\in [\tilde{\epsilon}+\Delta /2, \tilde{\epsilon}/4+\sqrt{(\Delta/2)^{2}_{}+(\epsilon^{*}_{+})^{2}_{}}]$, \\
3) $[1,0]$ for $\epsilon\in [\max[\tilde{\epsilon}-\Delta /2, \Delta/2], \tilde{\epsilon}+\Delta /2]$, \\
4) $[1,3]$ for $\epsilon\in [\min[\tilde{\epsilon}-\Delta /2, \Delta/2], \max[\tilde{\epsilon}-\Delta /2, \Delta/2]]$ if $\Delta <\tilde{\epsilon}$, and $[0,0]$ if $\Delta >\tilde{\epsilon}$, \\
5) $[0,3]$ for $\epsilon\in [-\Delta/2, \min[\tilde{\epsilon}-\Delta /2, \Delta/2]]$, \\
6) $[0,4]$ for $\epsilon\in [\tilde{\epsilon}/4-\sqrt{(\Delta/2)^{2}_{}+(\epsilon^{*}_{+})^{2}_{}},-\Delta /2]$ and \\
7) $[0,1]$ for $\epsilon_{{\rm F}}^{} < \tilde{\epsilon}/4-\sqrt{(\Delta/2)^{2}_{}+(\epsilon^{*}_{+})^{2}_{}}$.\\
These regions are highlighted in Fig.\ \ref{NoStrain}c, \ref{NoStrain}e and \ref{NoStrain}f.

iv) Finally, we can analyse the general case in which $v_{3}^{}$ and $v_{4}^{}$ are present in Eq.\ (\ref{Hlow}), with $v_{1}^{}\neq v_{2}^{}$. This can be of relevance for bilayer samples on a substrate, as the latter may induce an asymmetry in the two intralayer hopping energies $t_{1}^{}$ and $t_{2}^{}$.
In this case, for $\Delta=0$, the low energy spectrum is given by
\begin{eqnarray}
&&\epsilon_{\pm}^{}=\frac{v_{4}^{}\left( v_{1}^{}+v_{2}^{}\right)}{\gamma}\left|p\right|^{2}_{}\\
&&\quad\pm\sqrt{\left(\frac{v_{4}^{}\left( v_{1}^{}-v_{2}^{}\right)}{\gamma}\right)^{2}_{}\left|p\right|^{4}_{}+\left| v_{3}^{}p^{\dagger}_{}+\frac{v_{1}^{}v_{2}^{}+v_{4}^{2}}{\gamma}\, p^{2}_{}\right|^{2}_{}}\; ,\nonumber 
\end{eqnarray}
as illustrated in Fig.\ \ref{NoStrain}d.
The two bands touch only at $p=0$, where a single massless Dirac point survives while the other three disappear. In the experimentally relevant regime $v_{1}^{}-v_{2}^{}\ll \sqrt{v_{1}^{}v_{2}^{}}\simeq v$ the spectrum shows the opening of 
a minigap $\tilde{\Delta}\simeq \tilde{\epsilon}\, \left|v_{1}^{}-v_{2}^{}\right|/v$ at each of the three satellite Dirac points. Notice that no interlayer term $\Delta$ is required in order to open these minigaps. As a consequence a new regime appears for $\tilde{\epsilon}-\tilde{\Delta}/2<\epsilon_{{\rm F}}^{} <\tilde{\epsilon}+\tilde{\Delta}/2$, characterised by a FS of type $[1,0]$ similar to the regime 3) in the above case iii).

\section{Band structure with deformations}
\label{sec:Strain}

In the presence of generic elastic deformations of the lattice, the induced fictitious gauge fields modify the electronic low-energy Hamiltonian as in Eq.\ (\ref{Hlowdef}). 
While our formalism allows us to treat arbitrary distortions, as a first application we consider specific static lattice deformations which lead to uniform fictitious gauge fields, in analogy with those recently considered in the literature. \cite{Korean,Falko11} These gauge fields induce a shift in the electronic momenta that results in significant modifications to the band structure. The massless Dirac cones at low energy drift with the deformations until they annihilate at a critical value of strain. Increasing the strain further leaves a low energy spectrum made of two massless Dirac cones only. The modification to the band structure changes the nature of the LT as well as its energy. Thus, controllable strain could be used to induce the LT at the Fermi level, with observable consequences in the electronic DOS and other physical characteristics. The deformations leading to uniform gauge fields are uniaxial strain along an arbitrary direction, a rigid shift of one layer with respect to the other as well as a pure shear deformation. The first two types of distortions have been considered recently, \cite{Korean,Falko11} although the qualitative consequences in the spectrum are essentially the same for generic configurations leading to uniform gauge fields. 

A uniaxial in-plane strain along the direction $\hat{\theta}=(\cos\theta, \sin\theta)$ is described by the vector ${\bf u}^{(A)}_{}=h^{(S)}_{}=h^{(A)}_{}=0$ and ${\bf u}^{(S)}_{}({\bf r})=\beta_{\|}^{} r_{\|}^{}\hat{\theta}+\beta_{\perp}^{} r_{\perp}^{} \hat{\theta}_{\perp}^{}$, with $r_{\|}^{}={\bf r}\cdot\hat{\theta}$, $r_{\perp}^{}={\bf r}\cdot\hat{\theta}_{\perp}^{}$ and $\hat{\theta}_{\perp}^{}=\hat{z}\times\hat{\theta}$. Here $\beta_{\|}^{}$ and $\beta_{\perp}^{}$ represent the values of the strain along the two principal directions. This distortion results in a uniform deformation potential $D^{(S)}_{}$ which is reabsorbed in a global shift of the zero energy, while $D^{(A)}_{}=0$. The corresponding gauge fields are given by
$F^{(+)}_{l =1,2}=3a/4 (\partial t_{l}^{}/\partial a)(\beta_{\|}^{} -\beta_{\perp}^{} )\exp [-i2\theta ]$, $F^{}_{\gamma}=0$, $F^{(+)}_{3}=3a^2_{}/4\tilde{c}(\partial \gamma^{}_{3}/\partial \tilde{c})(\beta_{\|}^{} -\beta_{\perp}^{} )\exp [-i2\theta]$ and $F^{(+)}_{4}=3a^2_{}/4\tilde{c}(\partial \gamma_{4}^{}/\partial \tilde{c})(\beta_{\|}^{} -\beta_{\perp}^{} )\exp [-i2\theta]$.

Similarly, a pure in-plane shear of amplitude $\zeta$ can be described by the deformation vector ${\bf u}^{(A)}_{}=h^{(S)}_{}=h^{(A)}_{}=0$ and ${\bf u}^{(S)}_{}({\bf r})=\zeta r_{\|}^{}\hat{\theta}_{\perp}^{}$, resulting in the gauge fields 
$F^{(+)}_{l =1,2}=3a/4 (\partial t_{l}^{}/\partial a)(-i\zeta)\exp [-i2\theta ]$, $F^{}_{\gamma}=0$, $F^{(+)}_{3}=3a^2_{}/4\tilde{c}(\partial \gamma^{}_{3}/\partial \tilde{c})(-i\zeta)\exp [-i2\theta]$ and $F^{(+)}_{4}=3a^2_{}/4\tilde{c}(\partial \gamma_{4}^{}/\partial \tilde{c})(-i\zeta)\exp [-i2\theta]$.

Finally, a shift of the second layer with respect to the first one by the amount $\xi a$ along $\hat{\theta}$ is described by the deformation vector ${\bf u}^{(S)}_{}={\bf u}^{(A)}_{}=\xi a\hat{\theta}$, $h^{(S)}_{}=h^{(A)}_{}=0$, resulting in the gauge fields 
$F^{(+)}_{l =1,2}=0$, $F^{}_{\gamma}=-2(\partial \gamma/\partial c)(\xi a)^{2}_{}/c$, $F^{(+)}_{3}=3a^2_{}/\tilde{c}(\partial \gamma^{}_{3}/\partial \tilde{c})(i\xi )\exp [i\theta]$ and $F^{(+)}_{4}=3a^2_{}/\tilde{c}(\partial \gamma_{4}^{}/\partial \tilde{c})(-i\xi )\exp [i\theta]$.

\begin{figure}[hb]
	\centering
		\includegraphics[width=1.0\columnwidth]{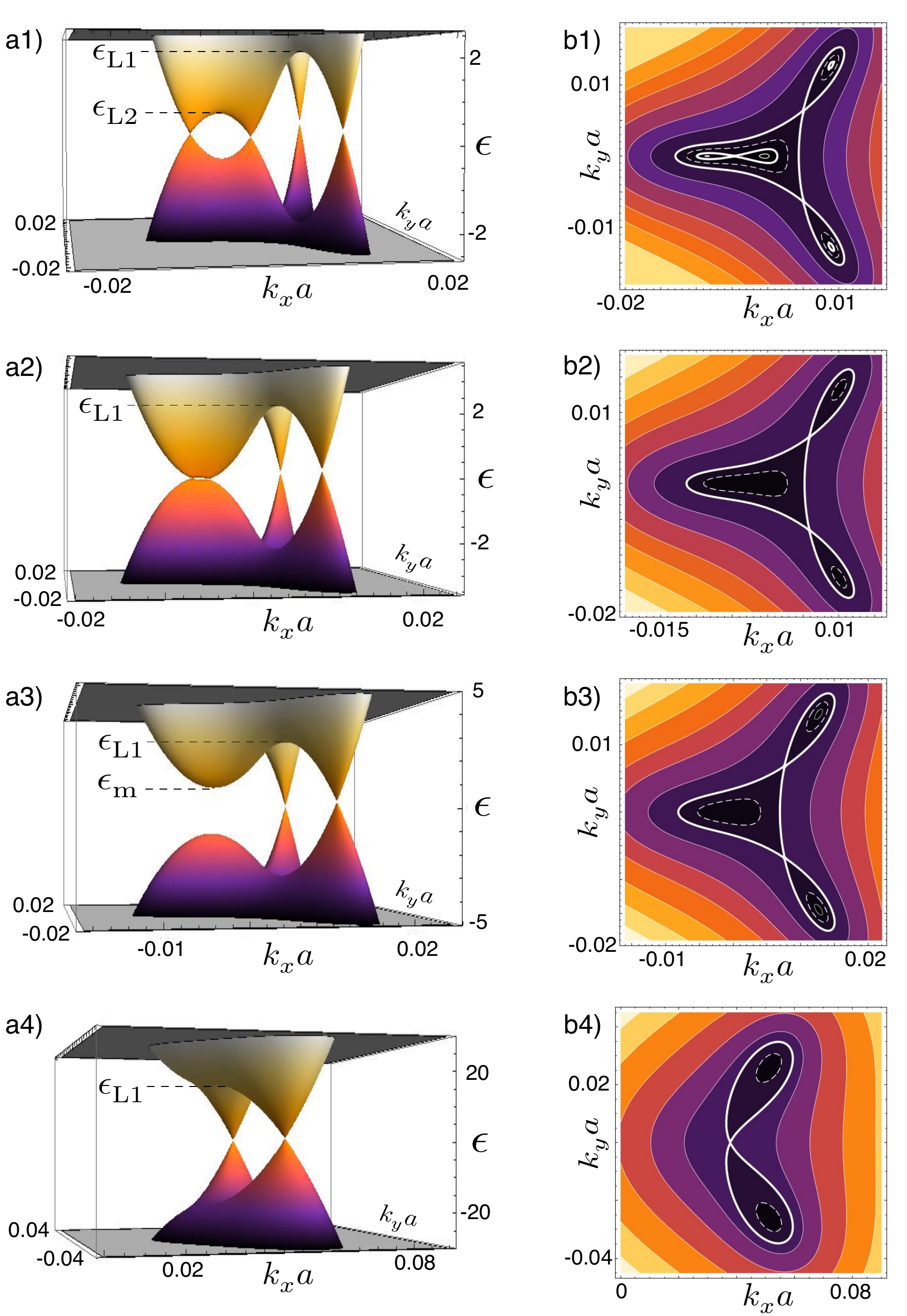}
			\caption{Electronic band structure in the wavevector space and equipotential lines for different values of strain $\beta$ along $\theta =0$, see text. Energy ($\epsilon$) is expressed in ${\rm meV}$ and the plots are taken for $t_{1}^{}=t_{2}^{}=2.47\, {\rm eV}$ and $\eta_{A1,B2}^{}=\eta_{A1,A2}^{}=\eta_{B1,B2}^{}=1$. For these parameters we get $\beta_{c1}^{} \simeq 2.3\cdot 10^{-3}_{}$ and $\beta_{c2}^{} \simeq 2\cdot 10^{-2}_{}$.
a1) Band structure for $\beta =1.5\cdot 10^{-3}_{}$. b1) Equipotential lines for a1), showing two LT at $\epsilon_{{\rm L1}}^{}$ and $\epsilon_{{\rm L2}}^{}$ (thick lines). For $\epsilon>\epsilon_{{\rm L1}}^{}$ the FS is of type [1,0] for this and all other panels. Dashed lines show the [3,0] FS at an energy between the two LT, while the thin lines exemplify a [4,0] FS at $\epsilon<\epsilon_{{\rm L2}}^{}$. a2) Band structure with critical strain $\beta_{c1}^{}$. b2) Equipotential lines for a2). One LT occurs at $\epsilon_{{\rm L1}}^{}$, below which the Fermi surface is of type [3,0]. a3) Band structure for $\beta =4\cdot 10^{-3}_{}$. The corresponding equipotential lines are shown in panel b3). The dashed line shows the [3,0] Fermi surface at $\epsilon_{{\rm m}}^{}<\epsilon<\epsilon_{{\rm L1}}^{}$, while thin lines show the [2,0] FS at $0<\epsilon<\epsilon_{{\rm m}}^{}$. a4) Band structure for $\beta =3\cdot 10^{-2}_{}$. The local minimum disappears. The corresponding equipotential lines are shown in panel b4). One LT occurs at $\epsilon_{{\rm L1}}^{}$. The dashed line shows a [2,0] FS for $0<\epsilon<\epsilon_{{\rm L1}}^{}$.
\label{Strain}}
\end{figure}

Quite generally, these different deformations translate into complex gauge fields in the Hamiltonian, which then affect the electronic band structure. The evolution of the electronic band structure under progressive strain is illustrated in Fig.\ \ref{Strain}.
Here we show the effect of a uniaxial strain of amplitude $\beta=\beta_{\|}^{}-\beta_{\perp}^{}$ along $\theta =0$, equivalent to a uniform shear of amplitude $\zeta =\beta$ along $\theta =\pi/4$. In Fig.\ \ref{Strain} we choose $t_{1}^{}=t_{2}^{}$ and $\eta_{A1,B2}^{}=\eta_{A1,A2}^{}=\eta_{B1,B2}^{}=1$ for illustration purposes. The electronic band structure in the wavevector space $(k_{x}^{},k_{y}^{})$ is shown in panels a1) to a4) at different values of $\beta$. While increasing $\beta$ two cones with chirality $\pi$ and $-\pi$ approach each other until they annihilate at a critical strain $\beta_{c1}^{}$. Increasing $\beta$ further induces the two fused cones to produce a local minimum at finite energy, until a second critical strain $\beta_{c2}^{}$ is reached. For $\beta >\beta_{c2}^{}$ the local minimum disappears, leaving two massless Dirac cones at low energy.
In Fig\ \ref{Strain}, panel a1) illustrates the band structure in the regime $0<\beta<\beta_{c1}^{}$ and b1) the corresponding equipotential lines for electronic states at positive energy. Two LT are visible at two different energies $\epsilon_{{\rm L1}}^{}> \epsilon^{*}_{}$ and $\epsilon_{{\rm L2}}^{}< \epsilon^{*}_{}$. The LT at $\epsilon_{{\rm L1}}^{}$ separates a FS of type [1,0] for $\epsilon>\epsilon_{{\rm L1}}^{}$ from a FS of type [3,0] for $\epsilon_{{\rm L2}}^{}<\epsilon<\epsilon_{{\rm L1}}^{}$. Similarly, for $0<\epsilon<\epsilon_{{\rm L2}}^{}$ the FS is of type [4,0]. Analogous results are obtained for hole doping at negative energies.\\
Panels a2) and b2) present the scenario for $\beta=\beta_{c1}^{}$. Two Dirac cones fuse at zero energy and only one LT is left at $\epsilon_{{\rm L1}}^{}$. Notice that the value of $\epsilon_{{\rm L1}}^{}$ grows while increasing the amount of strain. The LT separates two FS of type [3,0] and [1,0] for $0<\epsilon<\epsilon_{{\rm L1}}^{}$ and $\epsilon>\epsilon_{{\rm L1}}^{}$, respectively.\\
Panels a3) and b3) illustrate the regime $\beta_{c1}^{}<\beta<\beta_{c2}^{}$ where a local minimum at finite energy $\epsilon_{{\rm m}}^{}$ survives. The LT separates two FS of type [3,0] and [1,0] for $\epsilon_{{\rm m}}^{}<\epsilon<\epsilon_{{\rm L1}}^{}$ and $\epsilon>\epsilon_{{\rm L1}}^{}$, respectively. A new regime with FS of type [2,0] appears for $0<\epsilon<\epsilon_{{\rm m}}^{}$.\\
Finally, for $\beta>\beta_{c2}^{}$ a single LT occurs separating FS of types [2,0] and [1,0] for $\epsilon<\epsilon_{{\rm L1}}^{}$ and $\epsilon>\epsilon_{{\rm L1}}^{}$, respectively. This is illustrated in panels a4) and b4). It has to be noticed that in this regime of strain, the value of $\epsilon_{{\rm L1}}^{}$ is significantly larger than $\epsilon^{*}_{}$. The ability to tune the energy of the LT allows one to explore it at different levels of doping and also to partially overcome the problems in resolution due to disorder.

The qualitative picture above is reproduced, essentially unaffected, once the strain is applied at different angles $\theta$.\cite{Falko11} The precise values of the critical strains $\beta_{c1}^{}$ and $\beta_{c2}^{}$ depend on $\theta$ and on the values of the parameters $\partial t_{sl,s'l'}^{}/\partial {\ell}_{sl,s'l'}^{}\simeq -\eta_{sl,s'l'}^{} t_{sl,s'l'}^{}/{\ell}_{sl,s'l'}^{}$. Since only the intralayer derivatives are known ($\eta_{A1,B1}^{}=\eta_{A2,B2}^{}=3$), it is not possible to give a quantitative estimate of the critical strains. However, as shown in Fig.\ \ref{Strain}, for $\eta_{A1,B2}^{}=\eta_{A1,A2}^{}=\eta_{B1,B2}^{}=1$ the typical order of magnitude for them is around $0.2\,\% - 2\,\%$. This is easily achieved in realistic suspended graphene samples. 

These parameters seem to suggest that the low energy spectrum in conventional suspended bilayer samples with high mobility is probably characterised by two massless cones per valley, instead of four. As a consequence, in the presence of an external magnetic field, the degeneracy of the LL at zero energy is expected to be eight, due to two massless cones and two spins in each valley.\cite{Falko11} This picture is compatible with the recently observed integer quantum Hall effect at filling factors $\nu=\pm 4$ in bilayer graphene at low magnetic fields. \cite{Martin,MartinPRL} A detailed analysis of uniaxially strained bilayer graphene in the presence of magnetic fields has been recently discussed in Ref.\ \onlinecite{Falko11}. 

\begin{figure}[ht]
	\centering
		\includegraphics[width=1.0\columnwidth]{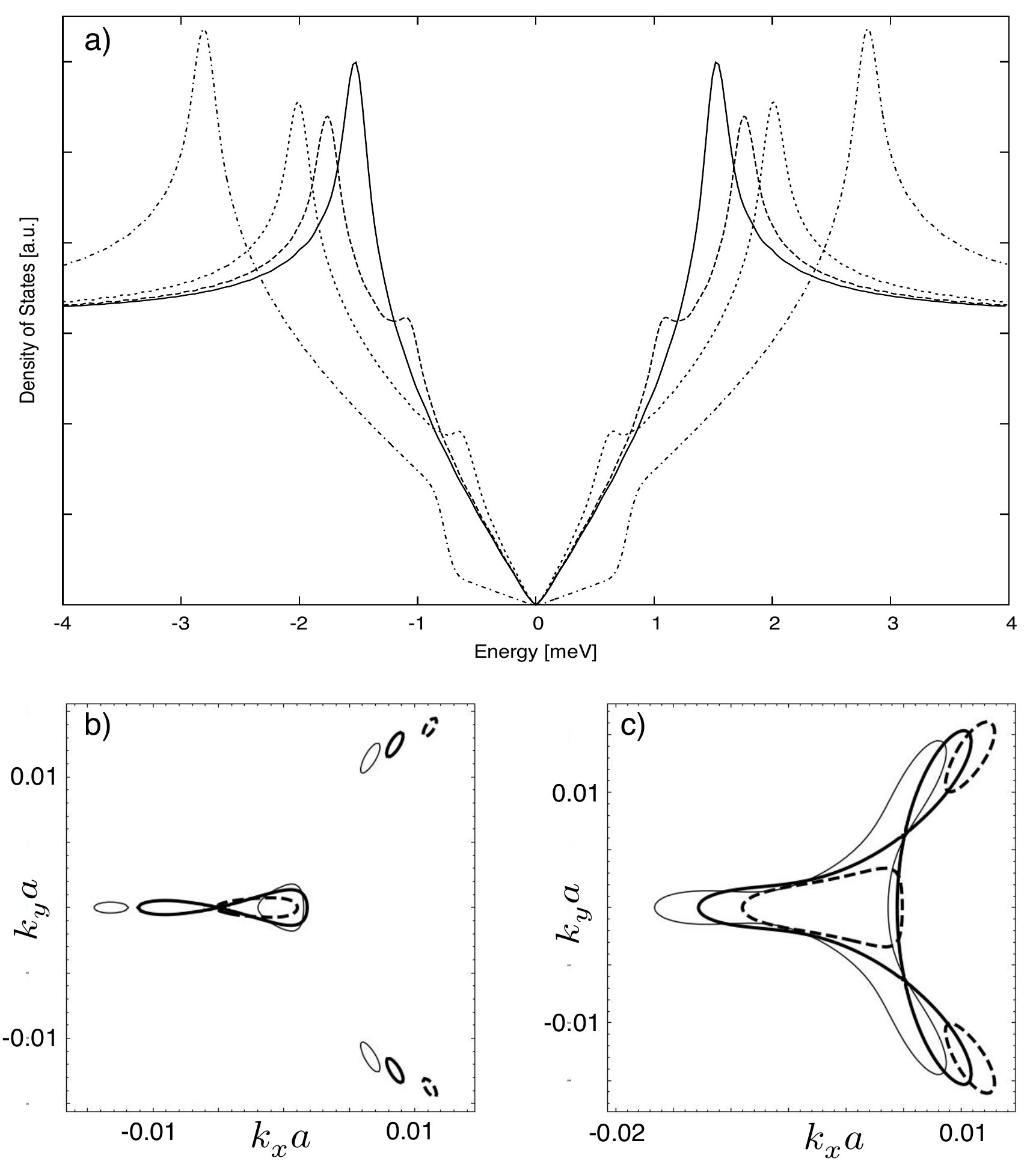}
			\caption{The electronic DOS and Fermi surface for different values of strain $\beta$ along $\theta =0$. Here we choose $\eta_{A1,B2}^{}=1$ and $\gamma_{4}^{}=0$. a) Electronic DOS as a function of energy for $\beta =0$ (thick line), $\beta =10^{-3}_{}$ (dashed line), $\beta =2\cdot 10^{-3}_{}$ (dotted line), $\beta =5\cdot 10^{-3}_{}$ (dot-dashed line). The peaks in the DOS at the various LT are clearly visible, as well as the linear dependence on energy in the low-energy regime due to the massless Dirac cones. The dot-dashed line shows a step-like feature at $\epsilon_{{\rm m}}^{}$ associated to the local parabolic minimum in the dispersion (see Fig.\ \ref{Strain}, panel a3). b) Fermi surface at electron doping corresponding to $\epsilon_{\rm F}^{}=0.8\, {\rm meV}<\epsilon^{*}_{}$. The thin line is for $\beta =0$, yielding a [4,0] FS. The thick line shows the FS at the LT ($\epsilon_{{\rm F}}^{}=\epsilon_{{\rm L2}}^{}$) for $\beta \simeq 1.2\cdot 10^{-3}_{}$, while the dashed line shows the [3,0] FS at $\beta \simeq 3\cdot 10^{-3}_{}$. c) Fermi surface at electron doping corresponding to $\epsilon_{\rm F}^{}=2.2\, {\rm meV}>\epsilon^{*}_{}$. The thin, thick and dashed lines correspond to $\beta=0$, $\beta \simeq 1.6\cdot 10^{-3}_{}$ and $\beta \simeq 3\cdot 10^{-3}_{}$, respectively. The LT here occurs at $\epsilon_{{\rm F}}^{}=\epsilon_{{\rm L1}}^{}$.\label{VaryingStrain}}
\end{figure}

As highlighted in Fig.\ \ref{Strain}, panels b1) to b4), the structure of the Fermi surface at a given density is thus affected by strain, due to the fusion of the Dirac cones. As a consequence, the nature of the Lifshitz transitions is sensitive to mechanical deformations. A controllable increase of the amount of strain in the bilayer graphene membrane could drive the Lifshitz transition at the Fermi level with observable consequences on the electronic DOS, as illustrated in Fig.\ \ref{VaryingStrain}. This effect could be directly detected in compressibility measurements as a function of density while keeping the strain constant or at a fixed density while tuning the strain. The modulation of the DOS at the Fermi level could also produce signatures in the linear conductivity in the absence of a magnetic field as long as the relevant diffusion coefficient is smooth across the LT.

Similar consequences to the strain-induced scenario described above have been found in terms of an interaction-induced spontaneous symmetry breaking leading to a nematic phase characterised by two massless Dirac cones at low-energy.\cite{Vafek,Falko10}. The latter scenario has been discussed in a very recent measurement on suspended bilayer graphene with extremely high mobility.\cite{Novoselov11}

\section{Phononic contribution to the resistivity via fictitious gauge field coupling}
\label{sec:Resistivity}

As a further application of the gauge fields we deduced in Sec.\ \ref{sec:Gauge}, we discuss the consequences of the corresponding electron-phonon coupling on the resistivity of suspended bilayer graphene. The interesting aspect to be pointed out is the appearance of a linear coupling between electrons and symmetric flexural phonons ($h^{(S)}_{}$) in the gauge field term $F_{3}^{(\tau)}$.\cite{NoteSmallF4} This is in contrast to the case of monolayer graphene, where flexural phonons have a quadratic coupling protected by symmetry with respect to the plane of the membrane. The contribution to the resistivity due to flexural modes is to be compared with the corresponding one due to in-plane phonons. 

In monolayers the competition is driven by a combination of dispersion and coupling of the various phononic modes. In-plane ones are hard to excite (they have a linear dispersion) but have a strong linear coupling to electrons. In parallel, flexural deformations are soft (they have a quadratic dispersion in the absence of tension) but have a weak quadratic coupling. If the tension is weak, it has been shown that flexural phonons dominate the in-plane modes as far as the resistivity is concerned.\cite{Mariani10,Castro} Recent measurements of the contribution to the resistivity due to electron-phonon scattering in suspended monolayers have indeed shown the dominant contribution by flexural phonons.\cite{Castro}

In bilayer graphene, the electron-phonon contribution to the resistivity has been considered recently in direct analogy with the monolayer case.\cite{Ochoa11,Heikkila,DasSarma11} In-plane phonons as well as flexural modes with a quadratic coupling stemming from intralayer deformations ($D^{(S)}_{}$, $F_{1}^{(\tau)}$ and $F_{2}^{(\tau)}$) preserve the qualitative temperature-dependence of monolayer samples. 
However, the appearance of a linear coupling for symmetric flexural modes offers the opportunity of investigating soft modes (for weak tension) with a strong linear coupling which were not considered so far.
Here we discuss this new contribution and analyse its consequences for transport.

\subsection{Phonon dispersion in bilayer graphene}

In order to study the electron-phonon contribution to the resistivity, we need to  discuss the dispersion of phonons in bilayer graphene.
This is easily obtained considering the elastic Lagrangian density \cite{LandauBook}
\begin{eqnarray}
&&{\cal L}=\sum_{l=1,2}^{}{\cal L}_{l}^{}+{\cal L}_{{\rm coupl}}^{}\quad\quad {\rm with}\nonumber\\
&&{\cal L}_{l}^{}=\frac{1}{2}\rho_{0}^{}\left(\dot{{\bf u}}^{(l)2}_{}+\dot{h}^{(l)2}_{}\right)-\frac{1}{2}\left(2\mu u^{(l)2}_{ij}+\lambda u^{(l)2}_{kk}\right)\nonumber\\
&&\quad\quad -\frac{1}{2}\left( \kappa(\nabla^{2}_{}h^{(l)}_{})^{2}_{}+\Gamma(\nabla h^{(l)}_{})^{2}_{}\right)\nonumber\\
&&{\cal L}_{{\rm coupl}}^{}=-\frac{1}{2}\rho_{0}^{}\Omega_{{\rm in}}^{2}\left({\bf u}^{(1)}_{}-{\bf u}^{(2)}_{}\right)^{2}_{}-\frac{1}{2}\rho_{0}^{}\Omega_{{\rm F}}^{2}\left(h^{(1)}_{}-h^{(2)}_{}\right)^{2}_{}\nonumber
\end{eqnarray}
where $\rho_{0}^{}$ is the mass density in each layer, $\mu$ and $\lambda$ are the Lam\'e coefficients for in-plane stretching, $\kappa$ is the bending energy and $\Gamma$ is a sample specific coefficient describing the degree of tension induced in the membrane.\cite{Parameters} The term ${\cal L}_{{\rm coupl}}^{}$ models a harmonic confinement for the sliding of one layer with respect to the other as well as for the modification of the interlayer distance. Recent first principle calculations \cite{Borysenko} produced the estimates $\Omega_{{\rm in}}^{}\simeq 5\cdot 10^{12}_{}\, {\rm Hz}$ and $\Omega_{{\rm F}}^{}\simeq 9\cdot 10^{12}_{}\, {\rm Hz}$. The Euler-Lagrange equations for the elastic Lagrangian at harmonic level are solved in terms of the symmetric and antisymmetric deformations ${\bf u}^{(\nu )}_{}({\bf r})=\sum_{{\bf q}}^{} {\bf u}^{(\nu )}_{{\bf q}}\exp [i{\bf q}\cdot {\bf r}]$ and $h^{(\nu)}_{}({\bf r})=\sum_{{\bf q}}^{} h^{(\nu )}_{{\bf q}}\exp [i{\bf q}\cdot {\bf r}]$, with $\nu =S,A$ and ${\bf u}^{(\nu )}_{{\bf q}}$, $h^{(\nu )}_{{\bf q}}$ their Fourier transforms in the wavevector space. The in-plane phononic eigenmodes are given by longitudinal and transverse components $u^{(\nu, L)}_{{\bf q}}={\bf u}^{(\nu )}_{{\bf q}}\cdot \hat{{\bf q}}$ and $u^{(\nu, T)}_{{\bf q}}={\bf u}^{(\nu )}_{{\bf q}}\cdot \hat{{\bf q}}_{\perp}^{}$, with $\hat{{\bf q}}={\bf q}/|{\bf q}|$ and $\hat{{\bf q}}_{\perp}^{}=\hat{z}\times\hat{{\bf q}}$. The dispersions of in-plane as well as flexural ($F$) deformations are 
\begin{eqnarray}
&&\omega^{(\nu, L)}_{{\bf q}}=\left[\frac{(2\mu +\lambda ) q^2_{}}{\rho_{0}^{}}+2\Omega_{{\rm in}}^{2}\, \delta_{\nu, A}^{}\right]^{1/2}_{}\nonumber\\
&&\omega^{(\nu, T)}_{{\bf q}}=\left[\frac{\mu q^2_{}}{\rho_{0}^{}}+2\Omega_{{\rm in}}^{2}\, \delta_{\nu, A}^{}\right]^{1/2}_{}\\
&&\omega^{(\nu, F)}_{{\bf q}}=\left[\frac{\kappa q^4_{}+\Gamma q^2_{}}{\rho_{0}^{}}+2\Omega_{{\rm F}}^{2}\, \delta_{\nu, A}^{}\right]^{1/2}_{}\; .\nonumber
\end{eqnarray}
These results hold at harmonic level, while anharmonic corrections due to the coupling between the two layers would stiffen the bending coefficient $\kappa$. In realistic bilayer graphene membranes symmetric flexural phonons would thus disperse as in the equation above, with a slightly renormalised bending energy. 
This is in analogy with the dispersion of bending modes in carbon-nanotubes.\cite{Mariani09}
The antisymmetric modes are gapped due to the harmonic interlayer couplings in ${\cal L}_{{\rm coupl}}^{}$ and do not give a relevant contribution to the resistivity in the linear transport regime. In the following we will thus concentrate on the symmetric flexural deformations with a linear coupling induced by $F_{3}^{(\tau)}$ and discuss their contribution in comparison with that of in-plane modes and of flexural modes with quadratic intralayer coupling discussed elsewhere.\cite{Ochoa11,Heikkila,DasSarma11} We will denote the dispersion of symmetric flexural modes as $\omega_{{\bf q}}^{}\equiv\omega_{{\bf q}}^{(S,F)}\sim \alpha q^2_{}$ (where $\alpha =\sqrt{\kappa /\rho_{0}^{}}\simeq 4.6\cdot 10^{-7}_{}\, m^{2}_{}/s$) for $q\gg q_{*}^{}$ and $\omega_{{\bf q}}^{}\sim \alpha q^{}_{*} q$ for $q\ll q_{*}^{}$, where $q_{*}^{}=\sqrt{\Gamma/\kappa}$ is a sample-specific wavevector related to the degree of external tension. Even in the absence of tension, anharmonic elastic corrections have been shown to induce a modification of the dispersion of flexural phonons at low energy.\cite{Mariani08,Nelson}

\subsection{The contribution of flexural phonons to the resistivity}

As far as the electron-phonon coupling is concerned, we consider the gauge field term coupling proportional to $F^{(\tau)}_{3}$ and focus on the high electron-density regime where the Fermi wave vector $k_{{\rm F}}^{}$ is larger than the inverse mean free path due to disorder and electron-phonon scattering. In this regime a quasiclassical Boltzmann approach to transport can be employed.\cite{Landau10} Except for extremely clean bilayer samples, this condition is fulfilled for $\epsilon_{{\rm F}}^{}>\epsilon^{*}_{}$, above the Lifshitz transition, where the effective electronic dispersion is parabolic. As a consequence, the relevant electronic and electron-phonon coupling Hamiltonians in one valley are given by
\begin{eqnarray}
&& H_{{\rm el}}^{(+)}\simeq \frac{1}{2m}\left(
\begin{array}{cc}
0 & p^{\dagger 2}_{}\\
p^{2}_{} & 0
\end{array}
\right)\nonumber \\
&& H_{{\rm el-ph}}^{(+)}\simeq \left(
\begin{array}{cc}
0 & F_{3}^{(+)}\\
F_{3}^{(+)\dagger} & 0
\end{array}
\right)\; .\nonumber
\end{eqnarray}
An electronic eigenstate with wavevector ${\bf k}$ and energy $\epsilon_{{\bf k}}^{}=\hbar^{2}_{}k^{2}_{}/2m$ is described by the spinor $|{\bf k}\rangle =1/\sqrt{2}\, (1, \exp [i2\phi_{{\bf k}}^{}])$, with $\phi_{{\bf k}}^{}$ the angle of ${\bf k}$ with respect to the ${\it x}$ axis.

In our case the dominant coupling due to symmetric flexural deformations is given by $F^{(+)}_{3}\simeq g^{}_{3}\, (\partial_{y}^{}h^{(S)}_{}-i\partial_{x}^{}h^{(S)}_{})$, with $g^{}_{3}=3ac/2\tilde{c}\, (\partial \gamma^{}_{3} /\partial \tilde{c})$ the coupling strength. In the Fourier space this corresponds to an electron-phonon coupling matrix
\begin{equation}
 \label{W}
w_{{\bf Q}}^{}= ig^{}_{3}Q\xi_{Q}^{} \left(\begin{array}{cc}0 & -i e^{i\Phi}_{}
\\ie^{-i\Phi}_{} & 0\end{array}\right) 
\end{equation}
in the Dirac description, where ${\bf Q}$ is the phonon wavevector, $\Phi$ its angle with respect to the $\hat{x}$ axis and $\xi_{Q}^{}=\left(\hbar /2M\omega_{{\bf Q}}^{}\right)^{1/2}_{}$ the oscillator length ($M$ the total oscillator mass per unit area).

In order to calculate the resistivity, a systematic derivation of the Boltzmann transport equation for electron-phonon coupling in graphene has already been presented in a previous work.\cite{Mariani10} The longitudinal resistivity is expressed as 
\begin{equation}
\label{Rho}
\rho =\frac{m}{ne^2 \tau_{k_{{\rm F}}^{}}^{}}
\end{equation}
where $n= k_{{\rm F}}^{2}/\pi$ is the electronic density and 
\begin{eqnarray}\label{Golden}
&&\frac{1}{\tau_{{\bf k}}^{}} =
-\frac{2\pi}{\hbar}\sum_{{\bf Q}}^{}2\omega_{{\bf Q}}^{}
\frac{\partial n_{{\bf Q}}^{}}{\partial \omega_{{\bf Q}}^{}}
\left(1-\cos\theta\right)\times \nonumber \\
&&\quad\quad\quad\quad\quad\quad\times\big|\langle {\bf k}+{\bf Q}|w^{}_{{\bf Q}}|{\bf k}\rangle\big|^{2}_{}
\delta (\epsilon_{{\bf k}+{\bf Q}}^{}-\epsilon_{{\bf k}}^{})
\end{eqnarray}
is the scattering rate at the Fermi level in the quasi-elastic approximation, due to both phonon absorption and emission processes. Here $\theta$ is the scattering angle between the electronic wave vectors ${\bf k}$ and ${\bf k}+{\bf Q}$ and $n_{{\bf Q}}^{}=1/(\exp [\hbar\omega_{{\bf Q}}^{}/k_{{\rm B}}^{}T]-1)$ is the equilibrium Bose distribution. The derivative of the Bose distribution implies that the relevant phonons to be considered have energies up to $\hbar\omega_{{\bf Q}}^{}\sim k_{{\rm B}}^{}T$ and their wavenumbers are restricted to $Q\lesssim q_{T}^{}$, with $\hbar\omega_{q_T^{}}^{}=k_{{\rm B}}^{}T$. In this regime $-\omega_{{\bf Q}}^{}
\partial n_{{\bf Q}}^{}/\partial \omega_{{\bf Q}}^{}\simeq k_{{\rm B}}^{}T/\hbar\omega_{{\bf Q}}^{}$.
Implementing the on-shell condition due to the quasielastic approximation we obtain 
\begin{eqnarray}
&&\big|\langle {\bf k}+{\bf Q}|w^{}_{{\bf Q}}|{\bf k}\rangle\big|^{2}_{}=g^{2}_{3}Q^{2}_{}\xi_{Q}^{2}\sin^{2}_{}(3\Phi)\; ,\nonumber \\
&& 1-\cos \theta =2 \cos^{2}_{}\Phi\nonumber
\end{eqnarray}
as well as 
\begin{equation}
\delta (\epsilon_{{\bf k}+{\bf Q}}^{}-\epsilon_{{\bf k}}^{})=\frac{2m}{\hbar^{2}_{}kQ\,\big|\sin\Phi_{0}^{}\big|}\delta(\Phi-\Phi_{0}^{})
\end{equation}
where $\Phi_0^{}$ is one of the two angles fulfilling the condition $\cos\Phi_{0}^{}=-Q/2k$. As a result, rescaling $Q$ by $2k_{{\rm F}}^{}$, we obtain
\begin{eqnarray}
&&\frac{1}{\tau_{k_{{\rm F}}^{}}^{}} = \frac{16\, m\, ng_{3}^{2}}{\hbar^{3}_{}\rho_{0}^{}}k_{{\rm B}}^{}T\, {\cal I}(k_{{\rm F}}^{},T)\quad ,\\
&&{\cal I}(k_{{\rm F}}^{},T)\simeq\int_{0}^{\min [\frac{q_{T}^{}}{2k_{{\rm F}}^{}},1]}{\rm d}x\, \frac{x^{4}_{}\left(1-4x^{2}_{}\right)^{2}_{}\sqrt{1-x^{2}_{}}}{\omega^{2}_{2k_{{\rm F}}^{}x}}\; .\nonumber
\end{eqnarray}
Different regimes appear as a function of temperature and of tension. The low temperature regime $q_{T}^{}/2k_{{\rm F}}^{}\ll 1$ corresponds to $T\ll T_{{\rm BG}}^{}$, with $T_{{\rm BG}}^{}=\hbar\omega_{2k_{{\rm F}}^{}}^{}/k_{{\rm B}}^{}$ the Bloch-Gr\"uneisen temperature, and is dominated by small angle scattering which give little contribution to the resistivity due to the term $1-\cos\theta$. In addition, for weak tension, flexural phonons in bilayer graphene are characterised by $T_{{\rm BG}}^{}\simeq 0.4\, \tilde{n}\, {\rm K}$, with $\tilde{n}$ the density expressed in units of $10^{12}_{}\, {\rm cm}^{-2}_{}$. As a consequence, for typical parameters of relevance to experiments, electron-phonon scattering yields a significant contribution to the resistivity only in the high-temperature regime, $T\gg T_{{\rm BG}}^{}$, where $q_{T}^{}\gg 2k_{{\rm F}}^{}$.
In this regime, for weak tension ($q^{}_{*}\ll 2k_{{\rm F}}^{}$) the resistivity takes the value
\begin{eqnarray}
&&\rho_{{\rm weak-tens}}^{(T\gg T_{{\rm BG}}^{})}\simeq \frac{h}{e^{2}_{}}\cdot \frac{m^{2}_{}g_{3}^{2}\, k_{{\rm B}}^{}T}{8\pi^{2}_{}\hbar^{4}_{}\kappa n^{2}_{}} \\
&&\quad\quad\simeq \frac{h}{e^{2}_{}}\cdot 10^{-4}_{}\, \frac{T[{\rm K}]}{\tilde{n}^{2}_{}}\simeq 2.6\, \frac{T[{\rm K}]}{\tilde{n}^{2}_{}}\, {\rm \Omega}\; .\nonumber
\end{eqnarray} 
In the opposite regime ($q^{}_{*}\gg 2k_{{\rm F}}^{}$) dominated by tension we obtain the sample specific result
\begin{equation}
\rho_{{\rm tens}}^{(T\gg T_{{\rm BG}}^{})} \simeq \frac{2k_{{\rm F}}^{2}}{q^{2}_{*}}\cdot \rho^{}_{{\rm weak-tens}}\ll \rho^{}_{{\rm weak-tens}}\; .
\end{equation} 
As in the case of monolayer graphene, external tension stiffens the flexural phonons and reduces their density of states without affecting their coupling, thereby suppressing the contribution to the resistivity.

For completeness, in the low temperature regime $T\ll T_{{\rm BG}}^{}$, we have the two results (for weak tension $q_{*}^{}\ll q_{T}^{}$ and strong tension $q_{*}^{}\gg q_{T}^{}$)
\begin{eqnarray}
&&\rho_{{\rm weak-tens}}^{(T\ll T_{{\rm BG}}^{})}\simeq \frac{h}{e^{2}_{}}\cdot \frac{8m^{2}_{}g_{3}^{2}\, (k_{{\rm B}}^{}T)^{3/2}_{}}{\pi\rho_{0}^{}\hbar^{2}_{}\left(\hbar\alpha (2k_{{\rm F}}^{})^{2}_{}\right)^{5/2}_{}}\\
&&\rho_{{\rm tens}}^{(T\ll T_{{\rm BG}}^{})}\simeq \frac{h}{e^{2}_{}}\cdot \frac{8m^{2}_{}g_{3}^{2}\, (k_{{\rm B}}^{}T)^{4}_{}}{3\pi\rho_{0}^{}\hbar^{2}_{}\left(\hbar\alpha q_{*}^{}(2k_{{\rm F}}^{})\right)^{5}_{}}\; . 
\end{eqnarray}   

As far as flexural phonons are concerned, the contribution to the resistivity we just considered has to be compared with the corresponding one stemming from the usual quadratic coupling (induced via the deformation potential $D^{(S)}_{}$ as well as the intra-layer gauge fields $F_{1}^{(\tau)}$ and $F_{2}^{(\tau)}$). In the absence of tension, and for $T\gg T_{{\rm BG}}^{}$, this has been estimated \cite{Ochoa11} to be 
\begin{equation}
\rho_{F}^{(T\gg T_{{\rm BG}}^{})}\simeq \frac{h}{e^{2}_{}}\cdot \frac{m^{2}_{}g_{F}^{2}\, (k_{{\rm B}}^{}T)^{2}_{}}{128\,\pi^{3}_{}\hbar^{4}_{}\kappa^{2}_{} n^{2}_{}}\; 
\end{equation} 
with $g_{F}^{}$ the screened deformation potential coupling constant. 
As a consequence, the ratio of the two contributions is
\begin{equation}
\frac{\rho_{{\rm weak-tens}}^{(T\gg T_{{\rm BG}}^{})}}{\rho_{F}^{(T\gg T_{{\rm BG}}^{})}}\simeq 5\cdot 10^{5}_{}\frac{g_{3}^{2}}{g_{F}^{2}\, T[{\rm K}]}\simeq \frac{500}{T[{\rm K}]}\; ,
\end{equation} 
where the last estimate is obtained with the approximations $g_{3}^{}\simeq -\gamma_{3}^{}/2\simeq 0.15\, {\rm eV}$ (corresponding to $\eta_{A1,B2}^{}\simeq 1$) and $g_{F}^{}\simeq 3.5\, {\rm eV}$. As a result, in the absence of tension and up to room temperature the linear coupling for flexural phonons mediated by the interlayer gauge fields yields a dominant contribution to the resistivity with respect to the quadratic one. In this case, the temperature dependent resistivity in suspended bilayer samples is expected to show a linear-$T$ dependence even in the absence of tension, in contrast to the monolayer case. Tension would then suppress the contribution due to flexural phonons in favour of the in-plane ones. 
The latter have been recently discussed in bilayer graphene.\cite{Ochoa11,Heikkila,DasSarma11} The resulting contribution to the resistivity is
\begin{equation}
\rho_{{\rm in}}^{(T\gg T_{{\rm BG}}^{})}\simeq \frac{h}{e^{2}_{}}\cdot 10^{-6}_{}\,\tilde{n}\,T[{\rm K}]\simeq 2.6\cdot 10^{-2}_{}\,\tilde{n}\,T[{\rm K}]\, {\rm \Omega}
\end{equation}
showing a linear-$T$ dependence as for $\rho_{{\rm weak-tens}}^{(T\gg T_{{\rm BG}}^{})}$ and $\rho_{{\rm tens}}^{(T\gg T_{{\rm BG}}^{})}$. 
The contribution due to in-plane phonons at room temperature is thus supposed to be about $10\, {\rm \Omega}$ at $\tilde{n}=1$, significantly less than flexural modes with weak tension. The critical tension $\Gamma$ needed to suppress $\rho^{(T\gg T_{{\rm BG}}^{})}_{{\rm tens}}$ with respect to $\rho_{{\rm in}}^{(T\gg T_{{\rm BG}}^{})}$ corresponds to an induced strain of order $3\cdot 10^{-3}_{}/\tilde{n}^{2}_{}$. Thus, as in monolayers, only samples with a small intrinsic strain have a chance to show signatures of flexural phonons.
Besides their magnitude, the in-plane and flexural contributions show the same temperature dependence, but behave differently with respect to density. Flexural modes would indeed yield a contribution to the resistivity which depends on the density as $n^{-2}_{}$ for weak tension and as $n^{-1}_{}$ in presence of strong tension. In contrast, in-plane modes result in a linear dependence of the resistivity on $n$. If experiments show a linear dependence of the electron-phonon resistivity on $T$, the corresponding density dependence would discriminate which phononic branch is dominant, and if tension is of special relevance.

\section{Conclusions}
\label{sec:Conclusions}

In this paper we analysed the consequences of generic elastic deformations on the electronic properties of bilayer graphene membranes. We deduced the fictitious gauge fields, induced by arbitrary distortions, in the electronic Dirac Hamiltonian and focused on the low-energy effective theory describing the two quasi-degenerate electronic bands close to zero energy.

As a first application we analysed the effect of static deformations yielding uniform fictitious gauge fields on the four massless Dirac cones induced by trigonal warping at low energy. An increasing degree of strain leads to the annihilation of two Dirac points, leaving the two remaining cones as the relevant low energy band-structure. This effect has direct consequences on the quantisation of the  Hall conductivity close to zero density.\cite{Falko11} In parallel, strain allows one to tune the topological Lifshitz transition at the Fermi level, with additional observable signatures in the single particle density of states.

We further considered the derived gauge fields as an electron-phonon coupling mechanism and discussed the consequent contribution to the temperature-dependent resistivity. We pointed out the appearance of a linear coupling between electrons and symmetric flexural phonons due to the inequivalence of the two layers and the non-vertical interlayer hopping processes. For suspended bilayer membranes with low tension, this results in a phononic contribution to the resistivity larger than the one stemming from the conventional quadratic coupling for flexural modes as well as that due to in-plane phonons.

Our investigation sets the basis for future works on the electromechanical properties of suspended bilayer graphene membranes, including applications in strain-engineering and in the creation of fictitious magnetic fields.

\begin{acknowledgments}
Useful discussions with Jens Martin, Saverio Russo and Guillaume Weick are gratefully acknowledged. FvO acknowledges financial support through SFB 658 and SPP 1459 of the Deutsche Forschungsgemeinschaft. 
\end{acknowledgments}


\begin{thebibliography}{99}
\bibitem{SchriefferWolfNote} This treatment is equivalent to the Schrieffer-Wolf transformation employed, e.g. in Ref.\ \onlinecite{Falko11}.
\bibitem{NoteSmallF4} The linear coupling for symmetric flexural modes appears also in the gauge field $F^{(\tau)}_{4}$. However, due to the smallness of the corresponding hopping parameter $\gamma_{4}^{}$, it leads to small corrections that will be neglected in our treatment of the resistivity.
\bibitem{Parameters} Typical parameters for graphene are $\kappa =1\,\mathrm{eV}$,  $\mu \simeq 4\lambda =9\,\mathrm{eV}{\rm \AA}^{-2}_{}$, $\rho_{0}^{}\simeq 7.6\cdot 10^{-7}_{}\, {\rm kg}\cdot {\rm m}^{-2}_{}$.
\end{thebibliography}
\end{document}